\documentclass[journal,twoside,web]{ieeecolorArxiv}
\usepackage{tmi_nologo}
\usepackage{tabularx}
\usepackage{amsmath,amssymb,amsfonts}
\usepackage{cite}
\usepackage{siunitx}
\usepackage[artemisia]{textgreek}
\usepackage{newunicodechar}

\usepackage[utf8]{inputenc}
\usepackage{color}
\usepackage{multirow}
\usepackage{chngcntr}

\usepackage{algorithm} 
\usepackage{algpseudocode} 
\usepackage{upgreek} 

\usepackage{graphicx}
\usepackage{textcomp}

\definecolor{magenta}{RGB}{139,0,139}
\newcommand{\Ab}{\mathbf{A}}
\newcommand{\Mb}{\mathbf{M}}
\newcommand{\xb}{\mathbf{x}}
\newcommand{\mb}{\mathbf{m}}
\newcommand{\bb}{\mathbf{b}}
\newcommand{\nb}{\mathbf{n}}
\newcommand{\zb}{\mathbf{z}}
\newcommand{\zbz}{\mathbf{z}^{(0)}}
\newcommand{\zbo}{\mathbf{z}^{(1)}}

\newcommand{\yb}{\mathbf{y}}
\newcommand{\vb}{\mathbf{v}}
\newcommand{\db}{\mathbf{d}}
\newcommand{\dbz}{\mathbf{d}^{(0)}}
\newcommand{\dbo}{\mathbf{d}^{(1)}}

\newcommand{\rFeat}{F_R}
\newcommand{\numIt}{N_{it}}
\newcommand{\rdnL}{m}
\newcommand{\rdnK}{l}

\usepackage{xcolor}
\definecolor{newcolor}{rgb}{.8,.349,.1}
 
\usepackage{url}
\usepackage{lipsum}
\usepackage{hyperref}
\hypersetup{
    colorlinks=true,
    linkcolor=blue,
    filecolor=magenta,      
    urlcolor=cyan,
    pdftitle={DEQ-MPI},
    }
\usepackage{gensymb}
\usepackage{caption}
\usepackage{comment}

\renewcommand{\arraystretch}{1.3}

\makeatletter
\IEEEtriggercmd{\reset@font\normalfont\fontsize{7.9pt}{8.40pt}\selectfont}
\makeatother
\IEEEtriggeratref{1}

\def\BibTeX{{\rm B\kern-.05em{\sc i\kern-.025em b}\kern-.08em
    T\kern-.1667em\lower.7ex\hbox{E}\kern-.125emX}}
{
}
\begin{document}
\title{DEQ-MPI: A Deep Equilibrium Reconstruction with Learned Consistency \\for Magnetic Particle Imaging}  
\author{Alper G\"{u}ng\"{o}r, Baris Askin, Damla Alptekin Soydan, Can Bar\i \c{s} Top, Emine Ulku Saritas,\\ Tolga \c{C}ukur, \vspace{-0.5cm} 
\\
 \thanks{This study was supported in part by TUBA GEBIP 2015 and BAGEP 2017 fellowships. The work of Alper G\"{u}ng\"{o}r was supported by TÜBİTAK BİDEB 2211 award. Corresponding author: A. G\"{u}ng\"{o}r (alperg@ee.bilkent.edu.tr).}
 \thanks{A. G\"{u}ng\"{o}r, E.U. Saritas, and T. \c{C}ukur are with the Department of Electrical and Electronics Engineering, and National Magnetic Resonance Research Center, Bilkent University, Ankara, Turkey (e-mails: \{alperg, saritas, cukur\}@ee.bilkent.edu.tr). A. G\"{u}ng\"{o}r is also with Aselsan Research Center, Ankara, Turkey. B. Askin is with Carnegie Mellon University, Pittsburgh, PA (baskin@andrew.cmu.edu). D.A. Soydan, and C.B. Top are with Aselsan Research Center, Ankara, Turkey (e-mails: \{dasoydan, cbtop\}@aselsan.com.tr).}
 }

\maketitle

\begin{abstract}
Magnetic particle imaging (MPI) offers unparalleled contrast and resolution for tracing magnetic nanoparticles. A common imaging procedure calibrates a system matrix (SM) that is used to reconstruct data from subsequent scans. The ill-posed reconstruction problem can be solved by simultaneously enforcing data consistency based on the SM and regularizing the solution based on an image prior. Traditional hand-crafted priors cannot capture the complex attributes of MPI images, whereas recent MPI methods based on learned priors can suffer from extensive inference times or limited generalization performance. Here, we introduce a novel physics-driven method for MPI reconstruction based on a deep equilibrium model with learned data consistency (DEQ-MPI). DEQ-MPI reconstructs images by augmenting neural networks into an iterative optimization, as inspired by unrolling methods in deep learning. Yet, conventional unrolling methods are computationally restricted to few iterations resulting in non-convergent solutions, and they use hand-crafted consistency measures that can yield suboptimal capture of the data distribution. DEQ-MPI instead trains an implicit mapping to maximize the quality of a convergent solution, and it incorporates a learned consistency measure to better account for the data distribution. Demonstrations on simulated and experimental data indicate that DEQ-MPI achieves superior image quality and competitive inference time to state-of-the-art MPI reconstruction methods.
\end{abstract}


\bstctlcite{IEEEexample:BSTcontrol}

\vspace{-0.2cm}
\section{Introduction}
\label{sec:introduction}
Magnetic particle imaging (MPI) is a powerful modality with high clinical prospect in applications such as angiography, cell tracking, cancer imaging, and neurovascular imaging \cite{Weizenecker_2009,zheng2015magnetic,arami2017tomographic,song2017janus,utkur2017,ludewig2017magnetic,cooley2018rodent,tay2018magnetic,tong2021highly}. MPI maps the spatial distribution of magnetic nanoparticles (MNPs) based on their magnetization responses \cite{gleich_tomographic_2005,Xspace_review}. A selection field (SF) creates a field free region for localized encoding, while a drive field (DF) evokes responses \cite{saritas2012rev}. The point spread function (PSF) in MPI is spatially variant and anisotropic due to system non-idealities (e.g., inhomogeneities in applied fields) and trajectory-dependent response of the MNPs. To account for these variations, a system matrix (SM) is typically utilized to characterize the PSF across the field-of-view (FOV) \cite{Knopp2010,GoodwillXspace2011, KilicSivp}. While analytical estimation is possible \cite{superResolutionMPISM}, experimentally measuring the SM with a calibration scan improves reliability against non-idealities \cite{ilbey2017comparison, gruttner2013formulation}. SM measurements are taken point by point, by traversing an MNP sample on a spatial grid covering the FOV at a desired resolution. Relatively compact grids are common in MPI given practical constraints on FOV and resolution due to hardware limitations (e.g., limited coil sensitivity, gradient strength), MNP properties (e.g., weak or wide responses, relaxation), and excessive calibration times (e.g., $\sim$12 hours for a 32$\times$32$\times$32 grid) \cite{von2017hybrid}.

Following calibration, an imaging scan is performed to map the MNP distribution in the anatomy of interest. For efficient encoding of the anatomical volume, field-free-line (FFL) scans can be performed by acquiring responses from an ensemble of MNPs located across a selected line \cite{cbtop2020}. By traversing the selected line along a trajectory, MNP responses can be acquired across the FOV. Since acquired data are linearly related to the MNP distribution via the SM, image reconstruction can be achieved by solving an inverse problem \cite{Knopp2010}. That said, MPI measurements carry significant correlations across the frequency dimension as the frequency response is governed by the MNP characteristics, and they are corrupted by high levels of correlated noise \cite{Paysen_2020}. These factors cause the SM to be rank deficient with respect to grid size, so the resultant inverse problem is underdetermined \cite{knoppIllPosedSVD}. As an underdetermined inverse problem, MPI reconstruction has high potential to benefit from regularization priors in order to recover high-quality images \cite{kamilovPnP,MoDL}. 

Traditional MPI reconstructions seek a solution that embodies both physical constraints related to the SM and acquired data, and attributes of high-quality images. While non-iterative solvers exist \cite{Takagi2015,Scmister2017}, optimization algorithms are prominent that iteratively enforce data consistency (DC) based on the SM and regularize the image \cite{ilbey2017comparison,Kluth2019}. For DC, hand-crafted measures based on energy or intensity differences between reconstructed and acquired data are common \cite{Knopp2010,Kluth2020}. For regularization, hand-crafted priors are used to promote desired attributes (e.g., smoothness, sparsity) via $\ell_2$ \cite{Knopp2010WeightedRecon,Konkle2015}, $\ell_1$ \cite{Lieb2021}, TV \cite{ilbey2017comparison} losses or their combinations \cite{Storath2017}. While they have been pervasive in MPI reconstruction, hand-crafted priors cannot fully capture the image distribution, they show suboptimal performance especially in regions with low signal (e.g., due to low coil sensitivity), and their performance depends on careful tuning of regularization weights that can vary substantially across scans \cite{Storath2017, deepImagePriorMPI_old}. 

In recent years, learning-based priors have received interest in MPI reconstruction as a powerful alternative. Purely data-driven methods train neural networks to directly recover images from frequency- or time-domain data \cite{nnReconChae,nnReconHatsuda,Koch2019,Gladiss1D}. Although they enable efficient inference, neglecting the physical constraints embodied in the SM can limit generalizability. To improve generalization, deep image prior (DIP) methods instead use untrained networks whose parameters are learned at test time by minimizing a DC loss \cite{deepImagePriorMPI_old,deepImagePriorMPI}. Yet, extensive inference times and difficulty in identifying appropriate architectures per dataset can limit utility \cite{knopp2022warmstart}. A recent plug-and-play method (PP-MPI) pre-trains an image prior for denoising and later combines it with the SM for reconstruction \cite{ppmpi}. The plug-ang-play framework offers a flexible compromise between efficiency and generalization in solution of inverse problems \cite{kamilovPnP, kamilovPnP2}. Yet, transferring a prior from the denoising to the reconstruction task can potentially elicit performance limitations. 

Here, we introduce a novel deep equilibrium model, DEQ-MPI, for improved performance and efficiency in MPI reconstruction. Inspired by physics-driven unrolling methods \cite{MoDL}, DEQ-MPI augments neural networks into an iterative optimization to rapidly alternate between regularization and DC projections. Conventional unrolled methods produce non-convergent solutions following a small number of iterations due to computational and memory constraints \cite{deq_original}, and they use hand-crafted DC measures that can elicit suboptimal performance \cite{Dong_liang}. DEQ-MPI instead trains an iterative architecture to maximize image quality at convergence for improved performance, and it introduces a novel learned consistency block to better conform to the MPI data distribution. Initialization strategies are also proposed for both regularization and learned consistency blocks to improve model training. Demonstrations show that DEQ-MPI achieves superior performance to state-of-the-art methods for MPI reconstruction, while also maintaining superior or on par efficiency. 

\vspace{-0.15cm} 
\subsection*{Contributions:}

\begin{itemize}
    \item We introduce the first physics-driven deep iterative architecture for performant and efficient MPI reconstruction.
    \item DEQ-MPI leverages the first deep equilibrium model and the first learned consistency measure in MPI.
     \item We propose initialization strategies for regularization and learned consistency blocks in DEQ-MPI to improve model training. 
\end{itemize}

\section{Related Work}
Learned image priors have recently been adopted as a promising approach in MPI tasks such as SM or image super-resolution \cite{2d-SMRnet,transmsTMI,Shang_2022}, view imputation in projection imaging \cite{PGNet}, and image reconstruction \cite{nnReconChae,nnReconHatsuda,Koch2019,Gladiss1D,deepImagePriorMPI,knopp2022warmstart,ppmpi}. For image reconstruction, purely data-driven methods provide fast inference by directly mapping acquired data onto images without explicitly considering the SM \cite{nnReconChae,nnReconHatsuda,Koch2019,Gladiss1D}. As these methods do not explicitly integrate physical constraints, reliability against system variability can be limited. Moreover, previous data-driven methods include dense layers whose complexity grows substantially with data dimensions. DIP methods instead use untrained networks with convolution filters serving as native regularizers, and learn network parameters to optimize DC on individual test scans \cite{deepImagePriorMPI,knopp2022warmstart}. Although DIP methods promise enhanced generalization by incorporating the SM, they require thousands of inference iterations and face challenges in network selection as ideal architectures are often image specific \cite{Arican2022}. PP-MPI pre-trains a convolutional network for image denoising, and combines it with the SM during an inference optimization \cite{ppmpi}. While PP-MPI offers improved efficiency compared to DIP, transferring a model from the denoising to the reconstruction task can limit performance \cite{adadiff}. Thus, learning-based methods with improved efficiency and generalization are needed in MPI reconstruction.
 
A powerful framework for learning-based reconstruction employs physics-driven unrolled methods that perform a fixed number of iterated projections through a convolutional network block to regularize the image and a DC block to enforce the system's physical constraints \cite{MoDL}. While no previous study has considered unrolled methods for MPI, state-of-the-art results have been reported with them in other modalities \cite{MoDL,Dar2020}. That said, as computational complexity grows rapidly when more blocks are cascaded, unrolled methods are typically trained to optimize image quality after a small number of iterations. This limitation results in suboptimal performance, and image quality degrades significantly when inference is sought at a different number of iterations than that prescribed for training as suggested by recent image reconstruction studies \cite{willett, dongLiangDEQPOCS, Heaton2021, deqMeasurementModel}. Moreover, DC in MPI and other modalities is commonly performed by projecting reconstructed data onto the $\ell_{2}$-ball of acquired data to alleviate bias due to noise \cite{Haldar2016,Knopp2010,Kluth2020,Lam2023}. This procedure ignores the underlying data distribution as it does not consider correlations among acquired data samples that can help lower such biases more effectively. In turn, the use of suboptimal DC measures can elicit performance losses during image reconstruction \cite{Dong_liang}.

Our proposed DEQ-MPI model leverages three technical novelties to address the limitations of conventional unrolled methods in the context of MPI reconstruction. First, DEQ-MPI is not trained to optimize performance within a fixed number of iterations, but rather upon convergence as inspired by recent deep equilibrium models in machine learning \cite{deq_original}. Second, DEQ-MPI introduces a novel learned consistency block based on a convolutional module as opposed to hand-crafted measures. Third, DEQ-MPI employs a novel initialization strategy for the learned consistency block to improve model training. DEQ-MPI introduces the first physics-driven iterative architecture, the first deep equilibrium model, and the first learned consistency measure for MPI in the literature. These technical advances enable DEQ-MPI to outperform state-of-the-art methods in MPI reconstruction. 

\begin{figure*}[t]
	\begin{minipage}{0.325\textwidth}
     \caption{\textbf{(a)} Conventional unrolled methods versus the deep equilibrium model in DEQ-MPI. Unrolled methods express reconstruction as repeated projections through a network operator, $\xb_{k+1}$=$h_\theta(\xb_k;\yb,\Ab)$ where $\xb_k$ is the image at iteration $k$, $\yb$ are acquired data, $\Ab$ is the system matrix, and network parameters $\theta$ are shared across iterations. $h_\theta$ is trained to optimize performance after $\numIt$ iterations where $\numIt$ is fixed and small to limit computational burden, resulting in suboptimal performance. DEQ-MPI instead leverages an implicit mapping $\xb_{*}$=$h_\theta(\xb_*;\yb,\Ab)$ to compute a convergent solution based on repeated injection of acquired data. In this case, $h_\theta$ is trained to maximize image quality upon convergence as opposed to an adhoc $\numIt$. \textbf{(b)} Proposed DEQ-MPI implementation. DEQ-MPI integrates the implicit mapping into an ADMM algorithm with fixed-point iterations expressed in Eq. \eqref{eq:fixedpointiter} for the image $\xb$ and Lagrange multipliers $\dbz, \dbo$. Each iteration involves projection through a learned regularization block ($\Psi_{RDN}$), projection through a learned consistency block ($\Psi_{LC}$), and reconciliation in a least-squares step to compute the output $\xb_{k+1}$.}
	\label{fig:methodMain}
	\end{minipage}
    \begin{minipage}{0.675\textwidth}
    \centering
	\centerline{\includegraphics[width=0.9\textwidth]{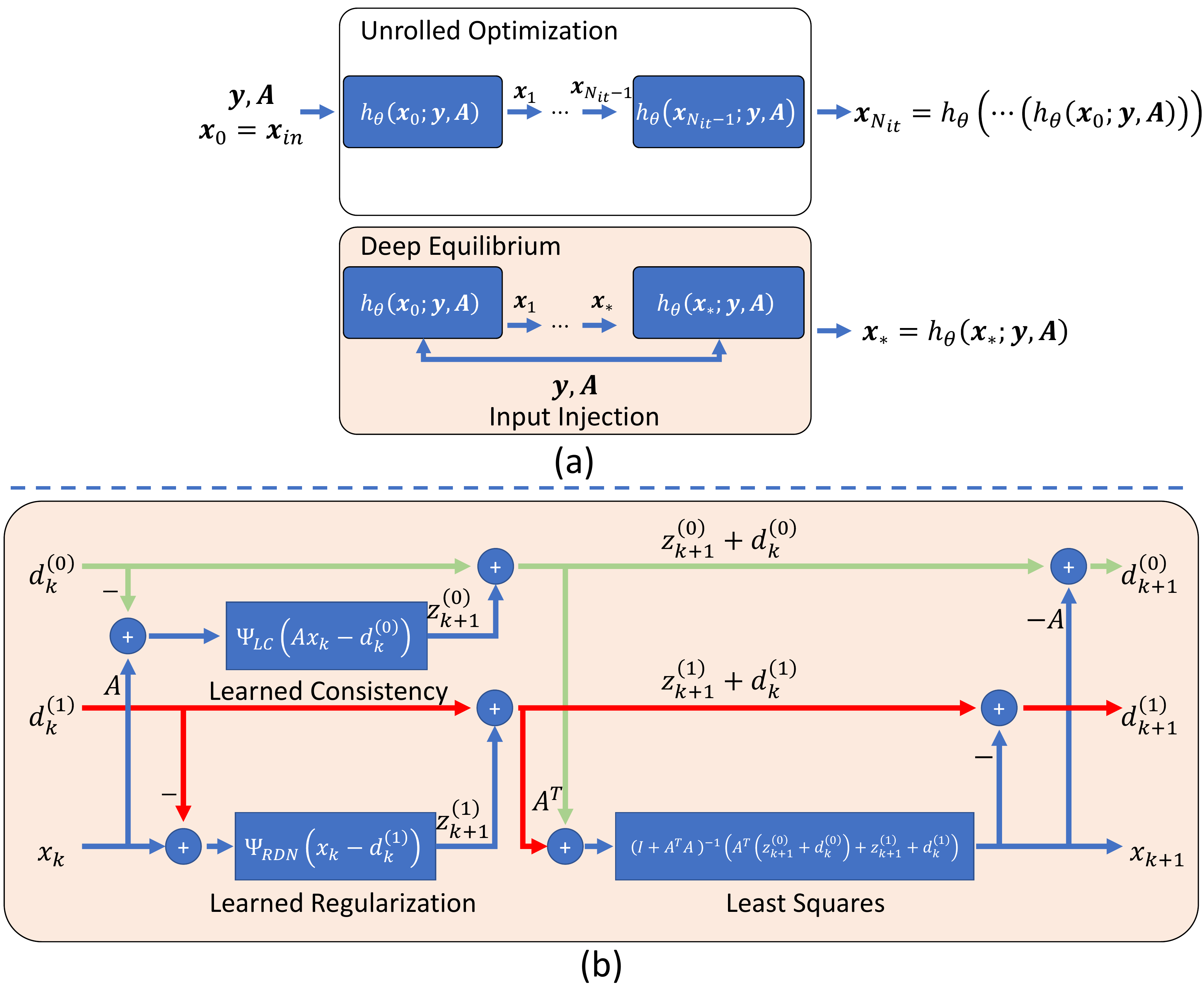}}
	\end{minipage}
\end{figure*}

\section{Theory}

\subsection{MPI Reconstruction}

Receive coils in MPI measure time-domain voltage waveforms that reflect the magnetization responses of MNPs. The acquired data can be transformed to frequency domain to define a linear system of equations \cite{gleich_tomographic_2005}:
\begin{align}
    \mathbf{Ax+n = y} \label{eq:mpiModel}
\end{align}
where $\mathbf{A} \in \mathbb{C}^{M \times N}$ is the SM, $\mathbf{x} \in \mathbb{R}^{N}$ is the image vector, $\mathbf{n} \in \mathbb{C}^{M}$ is the noise vector, $\mathbf{y} \in \mathbb{C}^{M}$ are frequency-domain data, $M$ is the number of frequency components, and $N$ is the number of voxels in the imaging grid. While $M$ is typically greater than $N$, both the SM $\mathbf{A}$ and the measurement noise $\mathbf{n}$ carry strong correlations across the frequency dimension \cite{Paysen_2020, cbtop2020}. As such, the inverse problem in Eq.~\eqref{eq:mpiModel} is underdetermined \cite{knopp_model-based_2010,knoppIllPosedSVD}. A common approach to solve Eq.~\eqref{eq:mpiModel} uses iterative optimization \cite{ilbey2017comparison,Kluth2019}:
\begin{align}
    \arg \min_{\xb\geq0} R(\xb)  \;\;s.t\;\; \Vert\Ab \xb - \yb\Vert_2<\epsilon, \label{eq:optimConst}
\end{align}
where $R(\cdot)$ is the regularization operator, $\epsilon$ is the error bound for the $\ell_2$-based DC measure that can be selected based on the estimated SNR \cite{cbtop2020, ilbey2017comparison}. Conventional methods adopt $R(\xb)=\sum_{i} \alpha_i r_i(\xb)$, where $r_i(\xb)$ is a hand-crafted function such as $\Vert \xb \Vert_2^2$, $\Vert \xb \Vert_1$, or $TV(\xb)$ \cite{Knopp2010WeightedRecon,Konkle2015,Lieb2021,Storath2017}.

An efficient algorithm is alternating direction method of multipliers (ADMM) that solves problems of type \cite{ilbey2017comparison}:
\begin{align}
    \arg\min_{\xb, \zb} g(\xb) + f(\zb) \text{ s.t. } \mathbf{Hx + Gz = c}, \label{eq:admm}
\end{align}
by splitting them into simpler sub-problems \cite{simitcs, gungorTCI22}. To arrive at an ADMM formulation equivalent to Eq. \eqref{eq:optimConst}, $\mathbf{H} = \left[ \Ab^T, \mathbf{I} \right]^T \in \mathbb{R}^{(M+N) \times N}, \mathbf{G} = -\mathbf{I} \in \mathbb{R}^{(M+N) \times (M+N)}, \mathbf{c} = 0$ with $\zb = \left[(\zbz)^T, (\zbo)^T\right]^T$ and $g(\xb) = 0$ can be selected:
\begin{align}
\arg\min_{\xb, \zb} f(\zb) \text{ s.t. } \xb = \zbo, \text{ and } \Ab\xb = \zbz , \label{eq:admmSolves}
\end{align}
where $\zb \in \mathbb{R}^{M + N}$ is the auxiliary variable vector used for splitting, and $f(\zb) = \chi(\zbz) + R(\zbo)$ where $\chi(\mathbf{t})$ is the indicator function of the set $\{ \mathbf{t}| \Vert \mathbf{t} - \yb \Vert_2\leq \epsilon \}$ for the DC constraint. The following iterations are used to solve Eq. \ref{eq:admmSolves}:
\begin{align}
\zbz_{k+1}&= \Psi_{\chi}(\Ab\xb_{k}-\dbz_k,\yb), \label{eq:sprob1} \\
\zbo_{k+1} &= \Psi_{R}(\xb_{k} - \dbo_{k}), \label{eq:sprob2} \\
\xb_{k+1} &= \Mb(\Ab^T(\zbz_{k+1}+\dbz_k)+\zbo_{k+1}+\dbo_k), \\
\dbz_{k+1} &= \dbz_{k}+\zbz_{k+1}-\Ab\xb_{k+1}, \\
\dbo_{k+1} &= \dbo_{k}+\zbo_{k+1}-\xb_{k+1}, 
\end{align}
where $k$ is the iteration index, $\Mb$=$(\mathbf{I}+\Ab^T\Ab)^{-1}$ can be precomputed, $\db$=$\left[(\dbz)^T, (\dbo)^T\right]^T \in \mathbb{R}^{M + N}$ contains Lagrange multiplier terms for the constraints in Eq. \ref{eq:admmSolves}. To incorporate the constraints flexibly, $\dbz$ captures data residuals due to deviation of $\Ab\xb$ from $\yb$ following DC, and $\dbo$ captures image residuals due to regularization of $\xb$. The proximal mappings for DC and regularization are given as:
\begin{align}
    \Psi_{\chi}(\vb,\yb) = \yb + \left\{ \begin{array}{cc} \vb - \yb & \mathrm{if} \mbox{ } \Vert \vb - \yb \Vert_2 \leq \epsilon \\
    \epsilon \frac{\vb - \yb}{\Vert \vb - \yb \Vert} & \mathrm{o.w.}
    \end{array}
    \right. , \label{eq:projl2} \\
    \Psi_{R}(\vb) = \arg\min_{\xb} R(\xb) + \frac{\mu}{2} \Vert \xb - \vb \Vert_2^2. \label{eq:proxR}
\end{align}
with $\mu$ scaled inversely with step size. Performance is limited by the capacity of $\Psi_{\chi}$ to describe MPI data distribution and the capacity of $\Psi_{R}$ to describe MPI image features.

\subsection{DEQ-MPI}
Unrolled methods use iterated projections through a network operator, $\xb_{k+1}$=$h_\theta(\xb_k;\yb,\Ab)$, with parameters $\theta$ commonly shared across iterations \cite{MoDL}. After a fixed number of iterations $\numIt$, $h_\theta$ is trained to optimize the quality of $\xb_{N_{it}}$:
\begin{align}
    \arg\min_{\theta} \left\Vert h_\theta(\dots h_\theta(h_\theta(\xb_0; \cdot);\cdot);\cdot) - \hat{\xb}_{r} \right\Vert_1 ,
\end{align}
where $\hat{\xb}_{r}$ is the ground truth image, and $\xb_0$ is an initial reconstruction estimate provided to the network at $k$=1 (Fig. \ref{fig:methodMain}a). Using large $\numIt$ improves performance by yielding solutions closer to the convergence point. Yet, while forward passes can be computed efficiently, backpropagation requires computation and storage of model gradients across all iterations, rendering large $\numIt$ prohibitive \cite{MoDL}. Thus, a small $\numIt$ is typically used that yields non-convergent solutions of limited quality. 

Unlike unrolled methods, DEQ-MPI leverages an implicit mapping $\xb_{*}$=$h_\theta(\xb_*;\yb,\Ab)$ based on a convergent solution $\xb_*$, as inspired by recent deep equilibrium models in machine learning \cite{deq_original}. In theory, an infinite number of iterations through $h_\theta$ might be required to obtain $\xb_*$. For efficiency, here we adopt an empirical convergence criterion to stop iterations when the relative change in $\xb$ between consecutive iterations falls below a small non-zero threshold \cite{deq_original,korkmaz2021unsupervised}. Training is then performed to maximize the quality of $\xb_*$ (Fig. \ref{fig:methodMain}a):
\begin{align}
    \arg\min_{\theta} \left\Vert h_\theta(\xb_*; \yb, \Ab) - \hat{\xb}_{r} \right\Vert_1,
    \label{eq:deq_loss}
\end{align}
Because a convergent solution is attained, DEQ-MPI can perform efficient backpropagation via implicit differentiation, where gradients have to be computed only at the convergent iteration for $\xb_*$ \cite{implicitLayersTutorial}. Since gradient terms for other iterations are not required, DEQ-MPI can improve performance without the computational overhead of unrolled methods.

Here we integrate the implicit mapping in DEQ-MPI into an ADMM algorithm with fixed-point iterations given as:
\begin{align}
\left[ \begin{array}{c} \xb \\
\dbz \\
\dbo \end{array} \right]_{k+1} &= 
h_\theta\left(\left[ \begin{array}{c} \xb \\
\dbz \\
\dbo \end{array} \right]_{k}; \yb, \Ab \right).
\label{eq:fixedpointiter}
\end{align}
Mapping through $h_{\theta}(\cdot)$ is then operationalized as (Fig. \ref{fig:methodMain}b):
\begin{align}
\zbz_{k+1}&= \Psi_{LC}(\Ab\xb_{k}-\dbz_k,\yb), \label{eq:DCFnc} \\
\zbo_{k+1} &= \Psi_{RDN}(\xb_{k} - \dbo_{k}), \label{eq:denoiserFnc} \\
\left[ \begin{array}{c} \xb \\
\dbz \\
\dbo \end{array} \right]_{k+1} &= 
\left[ \begin{array}{c} \Mb(\Ab^T(\zbz_{k+1}+\dbz_k)+\zbo_{k+1}+\dbo_k) \\
\dbz_{k}+\zbz_{k+1}-\Ab\xb_{k+1} \\
\dbo_{k}+\zbo_{k+1}-\xb_{k+1} \end{array} \right].
\end{align}
A solution for convergent $[\xb^T, \db^{(0),T}, \db^{(1),T}]_*^T$ is computed via fixed-point iterations accelerated with Anderson's method for efficiency \cite{anderson}. During these iterations, the proximal mapping $\Psi_{RDN}(\cdot)$ is implemented as projection through a residual dense network (RDN) block, where $\dbo$ captures image residuals after regularization. The proximal mapping $\Psi_{LC}(\cdot)$ is implemented as projection through a novel learned consistency (LC) block, where $\dbz$ captures data residuals after enforcement of consistency. As deep equilibrium methods can be sensitive to model initialization, we also introduce initialization strategies for both blocks. Details of model architecture and training procedures are discussed below.

\vspace{0.2cm}
\subsubsection{Model Architecture}
\textbf{RDN block: } $\Psi_{RDN}(\cdot)$ projects its input through a cascade of residual dense modules \cite{rdn}. The input in 2D form $\mathbf{v}=\xb_k-\db_k^{(1)} \in \mathbb{R}^{H\times W}$, where $\db_k^{(1)}$ captures image residuals, and $H$, $W$ are image height and width, passes through two convolutional layers $Z_{0}(\cdot)$:
\begin{align}
    \mathbf{u}_{0} = Z_{0}(\mathbf{v}).
\end{align}
The feature map $\mathbf{u}_0 \in \mathbb{R}^{\rFeat \times H \times W}$, where $\rFeat$ is the number of channels, is then processed with $n_{res}$ residual modules:
\begin{align}
    \mathbf{u}_{\rdnL} &= Z_{\rdnL}(\mathbf{u}_{\rdnL-1}),
\end{align}
where $Z_{\rdnL}(\cdot)$ is the $\rdnL^{\text{th}}$ module with $n_{conv}$ convolutional layers that receive concatenated outputs from previous layers:
\begin{align}
    \mathbf{u}_{\rdnL,\rdnK} &= Z_{\rdnL,\rdnK}([\mathbf{u}_{\rdnL-1}; \mathbf{u}_{\rdnL,1}; \mathbf{u}_{\rdnL,2}; \cdots; \mathbf{u}_{\rdnL,\rdnK-1}]),
\end{align}
where $\mathbf{u}_{\rdnL,\rdnK}\in\mathbb{R}^{F_S\times H \times W}$ is the output of $\rdnK^\text{th}$ convolutional layer, $Z_{\rdnL,\rdnK}$ with $1 \leq \rdnK \leq n_{conv}$. The output of $\rdnL^\text{th}$ residual module $\mathbf{u}_m$ is then computed by adding the module input to the output of a final convolutional layer, $Z_{\rdnL,out}$: 
\begin{align}
    \mathbf{u}_\rdnL &= Z_{\rdnL,out}([\mathbf{u}_{\rdnL-1}; \mathbf{u}_{\rdnL,1}; \cdots; \mathbf{u}_{\rdnL,n_{conv}}]) + \mathbf{u}_{\rdnL-1}.
\end{align}
The outputs of all residual modules are fused via a $1\times 1$ convolutional layer, $Z_{fuse}$:
\begin{align}
    \mathbf{u}_{fuse} &= Z_{fuse}([\mathbf{u}_{1}; \mathbf{u}_{2}; \cdots; \mathbf{u}_{n_{res}}]),
\end{align}
where $\mathbf{u}_{fuse}\in\mathbb{R}^{\rFeat\times H \times W}$. The output image $\mathbf{z}_{k+1}^{(1)}\in \mathbb{R}^{H\times W}$ is computed by a convolutional layer, $Z_{out}$, with ReLU activation to integrate a non-negativity constraint for MPI:
\begin{align}
    \mathbf{z}_{k+1}^{(1)} &= \text{ReLU}(Z_{out}(\mathbf{u}_{fuse}) + \vb)
\end{align}

\textbf{LC block: } A common approach to implement $\Psi_{\chi}(\cdot)$ in Eq. \eqref{eq:sprob1} is to project onto the $\ell_2$-ball of acquired data $\yb$. In contrast, DEQ-MPI leverages the LC block based on a convolutional module to better account for the MPI data distribution. LC receives a frequency-domain input $\mathbf{v} = \Ab \xb_k - \db_k^{(0)} \in \mathbb{C}^{M}$, where $\db_k^{(0)}$ captures data residuals, along with $\yb$:
\begin{align}
    \mathbf{z}_{k+1}^{(0)} = \Psi_{LC}(\mathbf{v, y}), 
\end{align}
where $\mathbf{z}_{k+1}^{(0)}$ are output data, and $LC$ is implemented as: %
\begin{align}
    \Psi_{LC}(\mathbf{v,y}) = \yb + \left\{ \begin{array}{cc} Z(\mathbf{v,y}) - \yb & \mbox{if } \Vert Z(\mathbf{v,y}) - \yb \Vert_2 \leq \epsilon \\
    \epsilon \frac{Z(\mathbf{v,y}) - \yb}{\Vert Z(\mathbf{v,y}) - \yb \Vert} & \mbox{o.w.}
    \end{array}
    \right. \label{eq:projLC}
\end{align}
In Eq. \eqref{eq:projLC}, $Z(\cdot)$ is a convolutional module with $n_{LC}$ hidden layers and $F_{LC}$ hidden units per layer. 
The $\epsilon$-bounded constraint prevents the output from diverging away from acquired data undesirably. Assuming field-free-line (FFL) scans with a single receive channel, data can be ordered in two dimensions as $\yb(f, \phi)$, where $f$ is the frequency component and $\phi$ is the FFL angle, and processed with 1D convolutional kernels across the frequency dimension. For field-free-point (FFP) scans with multiple receive channels, 2D kernels may instead be used over frequency and receive channel dimensions.

\subsubsection{Training Procedures}

\textbf{Model initialization:} 
Multiple convergent outputs $\xb_*$ can exist for the implicit mapping in DEQ-MPI, and the quality of a particular solution depends on the initialization of model parameters. Here, we propose to initialize the RDN block based on a plug-and-play approach as inspired by \cite{willett}. To do this, independent identically distributed (IID) Gaussian noise $\nb_{1} \in \mathbb{R}^{N}$ with standard deviation $\sigma_1$ is added onto a training set of MPI images $\hat{\xb}_r$, to generate images $\xb_{n} = \hat{\xb}_r + \nb_{1}$. RDN is pre-trained to suppress the additive noise in $\xb_{n}$:
\begin{align}
    \arg\min_{\theta_{RDN}} \left\Vert \Psi_{RDN}(\xb_{n}) - \hat{\xb}_r \right\Vert_1. \label{eq:rdnPretraining}
\end{align}
For the LC block, we propose a novel initialization procedure based on noise-added MPI data. First, noise-free data are generated using the SM and training MPI images, $\hat{\yb}_{r} = \Ab\hat{\xb}_r$. IID Gaussian noise is added at $\sigma_2$ and $\sigma_3$ to generate $\yb_n = \hat{\yb}_{r} + \nb_2$ and $\vb_n = \hat{\yb}_{r} + \nb_3$ with $\nb_{2,3} \in \mathbb{C}^M$, respectively. LC is pre-trained to mimic a canonical unlearned DC block: 
\begin{align}
    \arg\min_{\theta_{LC}} \left\Vert \Psi_{LC}(\vb_n,\yb_n) - \Psi_{\chi}(\vb_n,\yb_n) \right\Vert_1, \label{eq:LCinit}
\end{align}
where $\Psi_{\chi}$ is implemented as in Eq. \eqref{eq:projl2} based on projections onto the $\ell_2$-ball. We observed that pre-training to align the outputs of $\Psi_{LC}$ and $\Psi_{\chi}$ improves performance over pre-training to strictly align the output of $\Psi_{LC}$ with $\hat{\yb}_{r}$. While RDN and LC are initialized with the pre-trained weights for $\Psi_{RDN}(\cdot)$ and $\Psi_{LC}(\cdot)$, $\xb$ is initialized with the least-squares solution $\xb_{LS} = \Ab^{\dagger} \yb$ based on the pseudo-inverse of the SM, and $\dbz, \dbo$ are initialized as zero vectors.

\vspace{-1px}
\textbf{Implicit differentiation: } In a forward pass, a convergent solution of $\xb_* = h_\theta(\xb_*; \yb, \Ab)$ is computed via fixed-point iterations in Eq. \eqref{eq:fixedpointiter} accelerated using Anderson's method \cite{anderson}. An empirical convergence criterion is set as the $\ell_2$-norm difference between consecutive iterations falling below 10$^{-4}$. Here, a maximum of 25 iterations were observed to be sufficient for reaching convergence. In a backward pass based on Eq. \eqref{eq:deq_loss}, the Jacobian of the convergent solution $ \partial \xb_* / \partial \theta$ is computed by differentiating the implicit mapping:
\begin{align}
    \frac{\partial \xb_*}{\partial \theta} = \frac{\partial h_\theta(\xb_*) }{\partial \xb_*} \frac{\partial \xb_*}{\partial \theta} + \frac{\partial h_\theta(\xb_*) }{\partial \theta},
\end{align}
where the arguments $\yb$, $\Ab$ are omitted for brevity. The following solution for the Jacobian $ \partial \xb_* / \partial \theta$ is then obtained:
\begin{align}
    \frac{\partial \xb_*}{\partial \theta} = \left( \mathbf{I} - \frac{\partial h_\theta(\xb_*) }{\partial \xb_*} \right)^{-1} \frac{\partial h_\theta(\xb_*) }{\partial \theta}. \label{eq:gradAutoDiff}
\end{align}
Automatic differentiation tools for backpropagation require multiplication of the Jacobian with an arbitrary vector $\bb$ \cite{implicitLayersTutorial}:
\begin{align}
    \left(\frac{\partial \xb_*}{\partial \theta}\right)^T\bb = \left(\frac{\partial h_\theta(\xb_*) }{\partial \theta}\right)^T \left( \mathbf{I} - \frac{\partial h_\theta(\xb_*) }{\partial \xb_*} \right)^{-T} \bb. \label{eq:backwardPass}
\end{align}
To solve Eq.~\eqref{eq:backwardPass}, an intermediate vector can be defined as:
\begin{align}
    \mathbf{s}_* = \left( \mathbf{I} - \partial h_\theta(\xb_*) / \partial \xb_* \right)^{-T} \bb, \label{eq:defineS}
\end{align}
where $\partial h_\theta(\xb_*) / \partial \xb_*$ can be computed trivially. Eq.~\eqref{eq:defineS} can be rearranged to compute $\mathbf{s}$ via fixed-point iterations \cite{implicitLayersTutorial}:
\begin{align}
    \mathbf{s}_{i+1} = \left(\partial h_\theta(\xb_*) / \partial \xb_* \right)^{T} \mathbf{s}_i + \bb. \label{eq:rearrangeS}
\end{align}
The expression in Eq.~\eqref{eq:backwardPass} can then be evaluated based on $\mathbf{s}_*$:
\begin{align}
    \left(\partial \xb_* / \partial \theta\right)^T\bb = \left(\partial h_\theta(\xb_* )  / \partial \theta\right)^T \mathbf{s}_*.
\end{align}
As such, the implicit mapping enables calculation of the Jacobian $ \partial \xb_* / \partial \theta$ in terms of $\partial h_\theta(\xb_*) / \partial \xb_*$ and $\partial h_\theta(\xb_*) / \partial \theta$.

\section{Methods}

\subsection{Competing Methods}
DEQ-MPI was demonstrated against state-of-the-art methods based on hand-crafted and learned priors. For each method, hyperparameters were selected based on peak SNR (pSNR) performance on a validation set. The number of inference iterations was selected according to the L-curve criterion to achieve a favorable trade-off between performance and computation time \cite{korkmaz2021unsupervised}. Methods were implemented in PyTorch on a Tesla V100 GPU. Except DIP, learning-based models were trained for 200 epochs via the ADAM optimizer ($\beta_1=0.9$, $\beta_2=0.999$). Code to implement DEQ-MPI is available at {\small \url{https://github.com/icon-lab/DEQ-MPI}}.

\textbf{DEQ-MPI:} Architectural parameters were taken as $n_{res} = 4, \rFeat = 12, n_{conv} = 12$ for the RDN block, $n_{LC} = 1, F_{LC} = 8$ for the LC block. Cross-validated parameters included a learning rate of $10^{-3}$, 25 inference iterations. $\epsilon = \sqrt{M}$ for data consistency, and $\sigma_1 = 0.1$, $\sigma_2 = 0.05$, $\sigma_3 = 0.02$ for model initialization were used. 

\textbf{$\ell_1$-ADMM, TV-ADMM, Hyb-ADMM}: Three ADMM-based methods were implemented with $\ell_1$, TV, and a hybrid $\ell_1$+TV regularizer in the image domain, respectively \cite{ilbey2017comparison}. For each variant and each measurement SNR level, $\mu$ was selected to ensure convergence. $\mu = 250$ for the $\ell_1$, $\mu = 50$ for the TV, and $\mu = 10$ for the hybrid variant were used. For the hybrid variant, $\alpha_{TV} = 1 - \alpha_1$ was prescribed, and cross-validated values were $\alpha_1 = 0.1$ for SNR$<$20, $\alpha_1 = 0.8$ for 20$\leq$SNR$<$30, $\alpha_1 = 0.9$ for 30$\leq$SNR. Cross-validated number of iterations were 200 for the $\ell_1$, and 100 for the TV and hybrid variants. $\epsilon = \sqrt{M}$ was used.

\textbf{$\ell_2$-ART}: Algebraic reconstruction technique (ART), i.e. the Kaczmarz method, with a Tikhonov regularizer was implemented \cite{Knopp2010WeightedRecon}. Cross-validated parameters included 10 inference iterations, and a regularization weight of $\lambda = 10$ for SNR$<$15, $\lambda = 1$ for 15$\leq$SNR$<$35, $\lambda = 0.1$ for 35$\leq$SNR.

\textbf{DIP:} The DIP method based on an untrained network was implemented as described in \cite{deepImagePriorMPI}, albeit a mean-square error loss was adopted as it was observed to yield higher performance on the analyzed data. Inference was performed with the ADAM optimizer. Cross-validated parameters were $10^{-3}$ learning rate and 20000 inference iterations.

\textbf{PP-MPI:} The PP-MPI method based on a denoising prior was implemented \cite{ppmpi}. The network architecture and loss function were adopted from \cite{ppmpi}. Cross-validated parameters were $10^{-3}$ learning rate, additive noise with a standard deviation of $0.1$, and 150 inference iterations.

\begin{figure}[t]
	\centering
	\includegraphics[height=32.5mm]{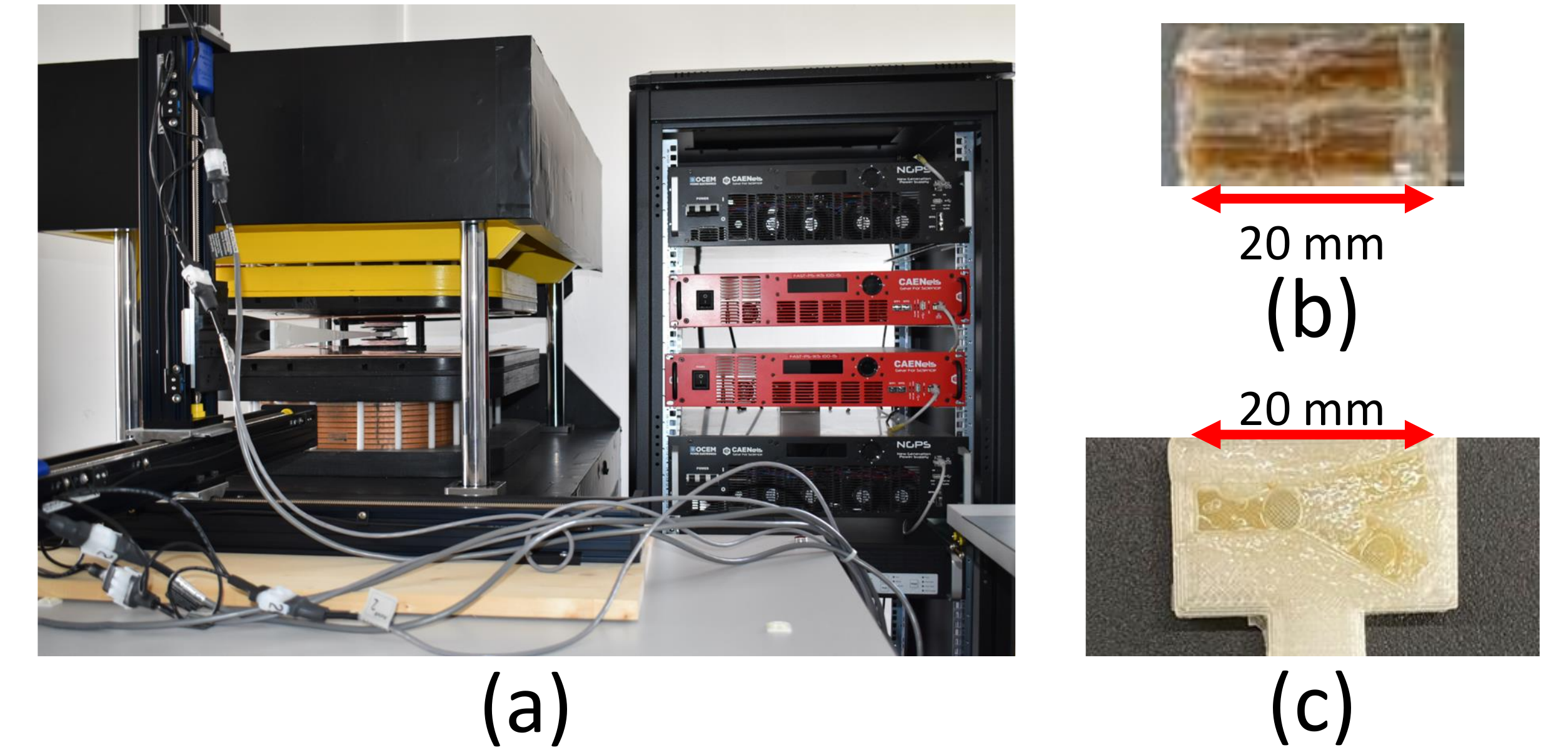}
	\caption{(a) Open-sided ASELSAN FFL scanner. (b) A cylindrical phantom with two parallel tubes was filled with Perimag MNPs at a dilution of 1:20. (c) A Y-shape phantom was filled with Perimag at a dilution of 1:100
  and contained a central air bubble.}
	\label{fig:scanner}
\end{figure}

\vspace{-0.2cm}
\subsection{MPI Phantoms}
\label{sec:phantoms}
Vessel phantoms are commonly used in demonstrating MPI reconstructions. Here, we generated simulated vessel phantoms based on time-of-flight magnetic resonance angiograms (MRA) \cite{ppmpi}. MRA images from 95 healthy subjects in the ITKTubeTK dataset were used \cite{URL_for_dataset}. Data were split into non-overlapping training, validation, test sets of 77, 9, 9 subjects, respectively. 10$\times$26$\times$52 volumetric patches were randomly cropped, followed by a maximum-intensity projection (MIP) along the first dimension, and downsampling in other dimensions onto 13$\times$26 images. The maximum pixel intensity in each image was randomly scaled to a number between 0.5 and 1.5. A total of 33692 training, 3377 validation and 3730 test images were obtained. Vessel phantoms were also generated at a larger grid size of 26$\times$52 following the same procedures, with the difference of starting from 10$\times$52$\times$104 volumetric patches.

We also generated simulated torus-shaped phantoms to systematically assess the resolvability of fine-grained image features. Three separate phantoms were generated with 4-mm tube diameter and 1, 2, or 3-mm inner torus diameters, corresponding to outer torus diameters of 5, 6, or 7 mm, respectively. The torus contained MNPs while the background was void. A continuous torus model was initially sampled at 0.1-mm resolution and then downsampled onto 1-mm resolution, resulting in $26\times52$ images. Noise was added to attain 15 dB measurement SNR. Multiple images were generated from each phantom by using 100 independent noise realizations.

For the experiments, two different phantoms were used (Fig. \ref{fig:scanner}b,c). The first included two parallel cylindrical tubes filled with Perimag (Micromod GmbH, Germany) MNPs at a dilution ratio of 1:20. Each tube had 20-mm length, 2-mm inner radius, 4-mm outer radius, and the tubes were attached together without a gap, so their center-to-center distance was 8 mm. The second included a 3D-printed `Y-shape' filled with Perimag MNPs at a dilution ratio of 1:100, and contained a central air bubble with 1.1-mm radius. Two arms of 10.7-mm length with 3.5-mm spacing at one end, and one arm of 8.8-mm length formed the Y-shape. All arms were 3.5-mm wide. 

\vspace{-0.2cm}
\subsection{Experimental Procedures}
Experimental SM and phantom measurements were performed on the open-sided ASELSAN FFL scanner (Fig. \ref{fig:scanner}a) \cite{cbtop2020}. For the SM acquisition, an undiluted Perimag sample of size 2$\times$2$\times$2~mm$^3$ was scanned with 2-mm steps while the FFL was rotated in the transverse plane over a 26$\times$52~mm$^2$ FOV. A DF of 9~mT amplitude and 10~ms duration per angle was applied at a 10\% duty cycle. An SR-560 pre-amplifier (SRS, MA, USA) amplified the signal at a gain of 5, filtered it at a frequency cut-off of 10-300~kHz, and the signal was then sampled at 5~MS/s. Frequency components around the $2^{nd}$-to-$11^{th}$ harmonics were selected over 500~Hz bandwidths. Whitening and background subtraction were performed based on background measurements. High-SNR rows were selected (SNR$>$5). Two separate experimental sessions were conducted. In a first session, SM and Y-shape phantom measurements were performed at an SF gradient strength of 0.5 T/m. In a second session that was held two months later, SM measurements were taken at SF gradient strengths of 0.3, 0.5 and 0.6 T/m, while cylindrical phantom measurements were taken at 0.5 and 0.6 T/m.

\begin{figure}[t]
\centering
\centerline{\includegraphics[width=\columnwidth]{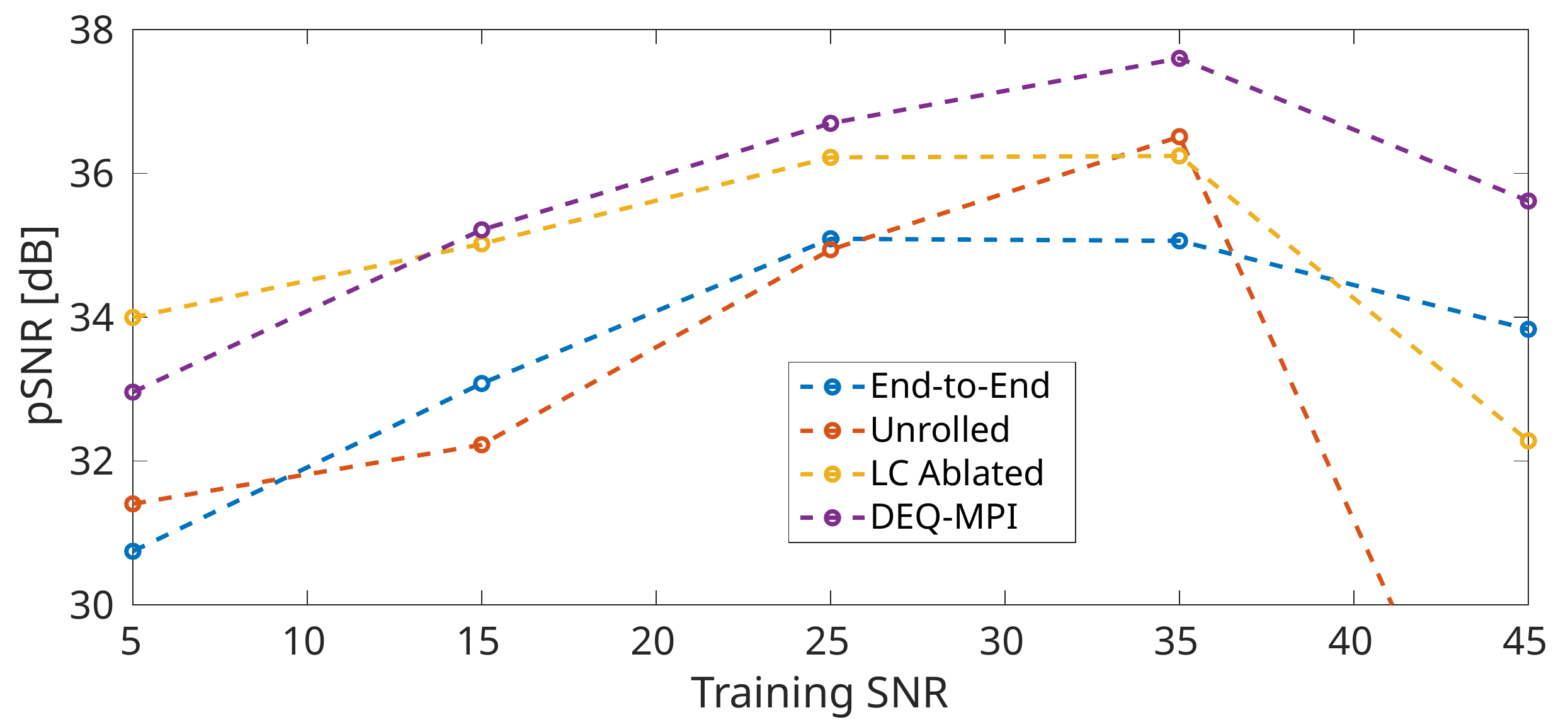}}
\caption{Comparison of DEQ-MPI against an end-to-end model that used an RDN block and omitted DC, an unrolled model with $\numIt$=5 iterations, and an LC-ablated variant based on an unlearned DC block. Separate models were trained at measurement SNRs of $5$-$45$~dB, testing was performed under $35$~dB SNR. Average pSNR across the test set is shown for each model.
}
	\label{fig:ablationGraph}
\end{figure}
\begin{figure}[t]
\centering
\centerline{\includegraphics[width=\columnwidth]{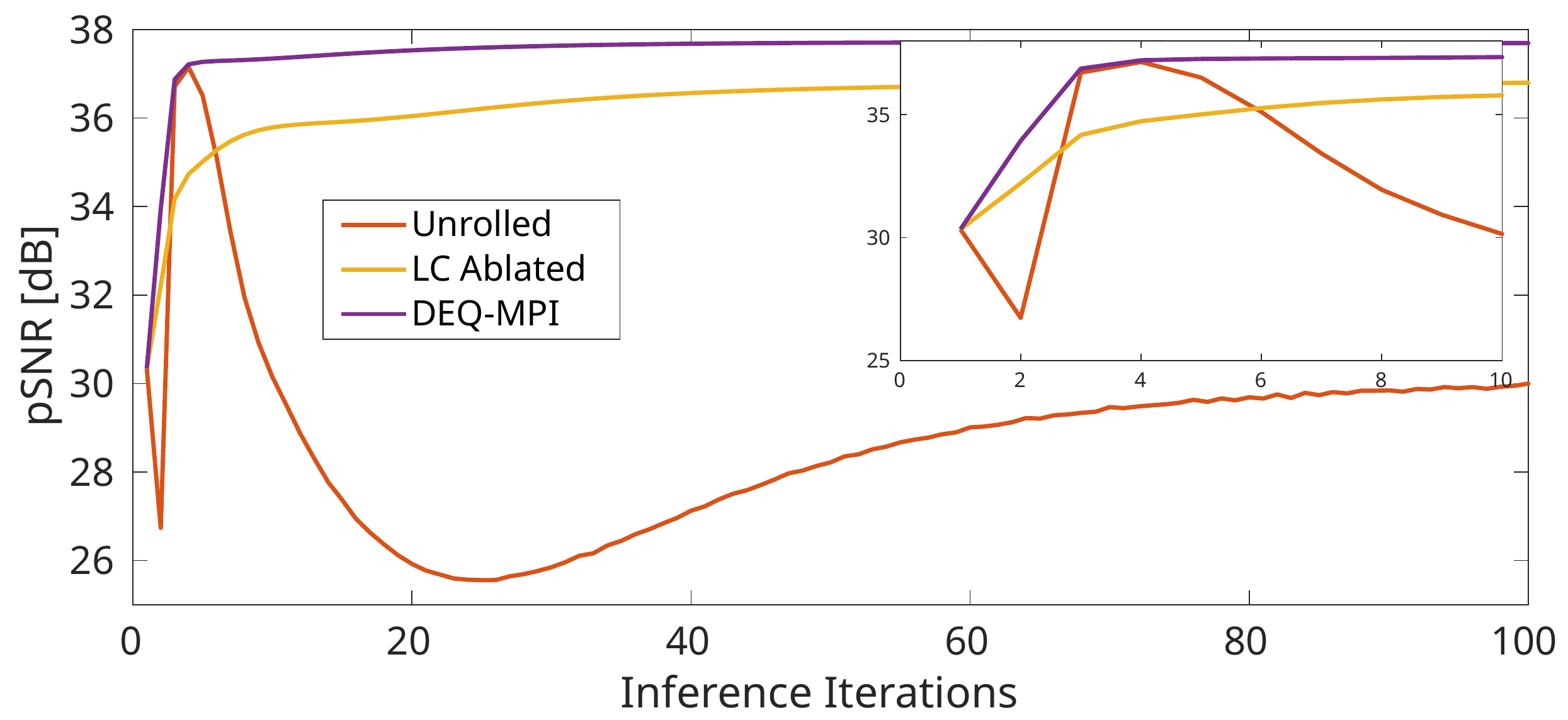}}
\caption{Convergence behaviors of DEQ-MPI, unrolled model, and LC-ablated variant. Models were trained and tested at $35$~dB SNR. Average pSNR across the test set is shown for each model. The zoom-in window highlights performance during initial iterations.
}
	\label{fig:ablationGraphConvergence}
\end{figure} 

\vspace{-0.2cm}
\subsection{Simulation and Analysis Procedures}
\label{sec:analysis}
For experimental phantoms, no ground-truth images exist to quantify reconstruction performance. Thus, to perform quantitative assessments in a setup that respects system non-idealities, we emulated MPI data by coupling simulated phantoms with experimental SMs. MPI measurements reflect a superposition integral between the continuously-varying system function and MNP distribution. Assuming that $\Ab_{c}$ and $\xb_{c}$ denote finely discretized SM and MNP distribution that closely approximate the underlying continuous variables: 
\begin{align}
\yb = \Ab_{c} \xb_{c} + \nb.
\end{align}
We do not have access to $\Ab_{c}$, but instead calibration scans capture SM on a discretized grid for a relatively large MNP sample size, $\Ab_{meas}=\Ab_{c} \mathbf{D}$ where $\mathbf{D} \in \mathbb{R}^{s^2N \times N}$ is the box-downsampling matrix by a factor of $s$ in two-dimensions \cite{transmsTMI}. Given $\Ab_{meas}$, one can approximate the system function by bicubic upsampling, $\tilde{\Ab}_{c} = \Ab_{meas} \mathbf{U}$ where $\mathbf{U} \in \mathbb{R}^{N \times s^2N}$. However, during reconstruction $\xb_{rec} = \mb(\yb, \tilde{\Ab}_{c})$, this elicits a discrepancy between the underlying SM that gives rise to the MPI data versus the estimated SM input to the reconstruction: 
\begin{align}
\xb_{rec} = \mb(\Ab_{c} \xb_{c} + \nb, \Ab_{c} \mathbf{D} \mathbf{U} ).
\label{eq:discrep}
\end{align}
Ignoring this discrepancy can lead to an inverse crime for simulation studies involving image reconstruction \cite{Bathke2017}. To avoid this problem, we generated MPI data by multiplying the simulated phantom with the SM measured during a calibration scan, i.e., $\hat{\yb}_{r} = \Ab_{meas}\hat{\xb}_r$. We then reconstructed the noise-added data $\yb = \hat{\yb}_{r} + \nb$  assuming a modified SM, $\Ab_{meas}\mathbf{U} \mathbf{D}$, to mimic the discrepancy highlighted in Eq.~\eqref{eq:discrep}.

For training DEQ-MPI, MPI data were generated by coupling simulated vessel phantoms from the training-validation sets with a single SM from the second experimental session. Separate models were trained for 13$\times$26 and 26$\times$52 phantoms, using the SM at SF gradient strength of 0.3, 0.5 or 0.6 T/m. To quantify model performance, simulated phantoms from the test set were coupled with the measured SMs from both experimental sessions. The DEQ-MPI models trained on simulated phantoms were also tested on experimental phantoms. During training and testing with 26$\times$52 phantoms, the measured SMs were upsampled by a factor of 2 via bicubic interpolation to have 1-mm/pixel resolution.

To describe the noise level in the generated MPI data, the measurement SNR was computed as:
\begin{align}
    SNR(\mathbf{y}) = 20 \log_{10} &\left(\Vert  \hat{\yb}_{r}\Vert_2 / \Vert  \yb - \hat{\yb}_{r}\Vert_2 \right),
\end{align}
where $\hat{\yb}_{r}$ and $\yb$ are noise-free and noisy data. To assess reconstruction performance for simulated phantoms, pSNR and structural similarity (SSIM) were computed:
\begin{eqnarray}
    && pSNR(\mathbf{x})= 20 \log_{10} \left(\sqrt{N} \Vert \hat{\mathbf{x}}_{r}\Vert_{\infty} / \Vert \mathbf{x}-\hat{\mathbf{x}}_{r}\Vert_2 \right), \\
    && SSIM(\mathbf{x})= \frac{(2 \mu_x \mu_{\hat{x}_r} + c_1)(2 \sigma_{x\hat{x}_r} + c_2)}{(\mu_x^2 + \mu_{\hat{x}_r}^2 + c_1)(\sigma_x^2 + \sigma_{\hat{x}_r}^2 + c_2)},
\end{eqnarray}
where $\xb$ is the reconstructed image, $\hat{\xb}_{r}$ is the ground truth, $\mu$, $\sigma^2$ denote image mean and variance, $\sigma_{x\hat{x}_r}$ is the covariance of $\xb$ and $\hat{\xb}_{r}$, $c_{1,2}$ are scalars that prevent division by zero \cite{SSIM_paper}.

\begin{table}[t]
\centering
\caption{Simulated vessel phantoms (13$\times$26) reconstructed at varying measurement SNRs. An SM at 0.5~T/m SF gradient was used in training and testing. pSNR (dB)\,/\,SSIM (\%) are reported as mean$\pm$std. across the test set. Boldface marks the top-performing method.} 
\label{table:seenDatasetResults}
\setlength{\tabcolsep}{1pt}
\renewcommand{\arraystretch}{1.5}
\resizebox{1.01\columnwidth}{!}{
\begin{tabular}{|c|c|c|c|} 
\hline 
 & SNR=$15$~dB & SNR=$25$~dB & SNR=$35$~dB  \\ 
 \hline 
\multirow{1}{*}{$\ell_1$-ADMM} & 18.9$\pm$3.5\,/\,16.4$\pm$9.5 & 23.2$\pm$2.2\,/\,32.6$\pm$14.0  & 28.9$\pm$1.3\,/\,58.6$\pm$13.8  \\ 
 \hline 
\multirow{1}{*}{TV-ADMM} & 27.3$\pm$6.1\,/\,56.2$\pm$20.0 & 31.2$\pm$4.4\,/\,71.9$\pm$11.3  & 35.5$\pm$3.0\,/\,84.5$\pm$6.5 \\ 
 \hline 
\multirow{1}{*}{Hyb-ADMM} & 28.7$\pm$5.1\,/\,63.9$\pm$9.5 & 31.7$\pm$3.6\,/\,74.8$\pm$7.5 & 35.9$\pm$2.7\,/\,85.6$\pm$5.6 \\ 
 \hline 
\multirow{1}{*}{$\ell_2$-ART} & 26.7$\pm$4.3\,/\,49.6$\pm$7.8 & 28.5$\pm$5.7\,/\,64.3$\pm$8.8 & 31.9$\pm$4.3\,/\,76.1$\pm$4.4 \\ 
 \hline 
\multirow{1}{*}{DIP} & 16.7$\pm$2.6\,/\,13.5$\pm$9.1 & 22.6$\pm$1.7\,/\,30.9$\pm$16.6 & 28.2$\pm$2.9\,/\,54.6$\pm$21.0 \\ 
 \hline 
\multirow{1}{*}{PP-MPI} &  30.6$\pm$5.1\,/\,70.8$\pm$11.5  & 33.4$\pm$4.1\,/\,77.8$\pm$9.6 & 35.7$\pm$3.1\,/\,82.8$\pm$8.2 \\ 
 \hline 
\multirow{1}{*}{DEQ-MPI} & \textbf{32.1$\pm$4.5}\,/\,\textbf{75.2$\pm$9.9} & \textbf{34.8$\pm$3.5}\,/\,\textbf{81.5$\pm$8.3}  & \textbf{37.7$\pm$2.6}\,/\,\textbf{88.1$\pm$5.7} \\ 
 \hline 
\end{tabular}}
\end{table}

\begin{table}
\centering
\caption{pSNR (dB)\,/\,SSIM (\%) of the simulated vessel phantoms (13$\times$26) reconstructed at varying measurement SNRs. The SM at 0.5 T/m was used in training, and the SM at 0.6 T/m was used in testing.} 
\label{table:unseenDatasetResults}
\setlength{\tabcolsep}{1pt}
\renewcommand{\arraystretch}{1.5}
\resizebox{1.01\columnwidth}{!}{
\begin{tabular}{|c|c|c|c|} 
\hline 
 & SNR=$15$~dB & SNR=$25$~dB & SNR=$35$~dB  \\ 
\hline
\multirow{1}{*}{$\ell_1$-ADMM} & 19.1$\pm$3.5\,/\,17.1$\pm$9.5  & 23.3$\pm$2.2\,/\,33.4$\pm$14.1  & 29.1$\pm$1.3\,/\,59.3$\pm$13.6  \\
 \hline 
\multirow{1}{*}{TV-ADMM} & 27.5$\pm$6.1\,/\,57.4$\pm$19.5  & 31.3$\pm$4.4\,/\,72.8$\pm$10.8  & 35.7$\pm$3.0\,/\,84.9$\pm$6.3  \\ 
 \hline 
\multirow{1}{*}{Hyb-ADMM} & 28.7$\pm$5.0\,/\,64.4$\pm$9.4  & 31.8$\pm$3.6\,/\,75.3$\pm$7.3  & 36.0$\pm$2.7\,/\,86.0$\pm$5.4  \\ 
 \hline 
\multirow{1}{*}{$\ell_2$-ART} & 26.7$\pm$4.3\,/\,49.8$\pm$8.0  & 28.5$\pm$5.8\,/\,64.6$\pm$8.8  & 32.0$\pm$4.3\,/\,76.5$\pm$4.4  \\ 
 \hline 
\multirow{1}{*}{DIP} & 16.9$\pm$2.6\,/\,13.9$\pm$9.0  & 22.5$\pm$1.8\,/\,31.0$\pm$16.7  & 28.0$\pm$3.1\,/\,54.1$\pm$21.8  \\ 
 \hline 
\multirow{1}{*}{PP-MPI} & 30.8$\pm$5.0\,/\,71.3$\pm$11.4  & 33.5$\pm$4.1\,/\,78.4$\pm$9.2  & 35.8$\pm$3.1\,/\,83.4$\pm$7.8  \\ 
 \hline 
\multirow{1}{*}{DEQ-MPI} & \textbf{31.9$\pm$4.5}\,/\,\textbf{75.0$\pm$9.8}  & \textbf{34.5$\pm$3.4}\,/\,\textbf{81.3$\pm$8.1}  & \textbf{37.6$\pm$2.6}\,/\,\textbf{88.0$\pm$5.7}  \\ 
 \hline 
\end{tabular}
}
\end{table}

\vspace{-0.05cm}
\section{Results}
\vspace{-0.2cm}
\subsection{Ablation Studies}
We conducted a set of ablation studies to assess the value of the individual design elements in DEQ-MPI. The ablation studies were conducted using 13$\times$26 simulated vessel phantoms and measured SMs. To assess the value of physics-driven learning, an end-to-end model was built where an RDN block without DC was trained to directly map the least-squares solution onto ground-truth images. To assess the value of deep equilibrium modeling, an unrolled model with conventional unlearned DC block was built with $\numIt=5$ iterations (selected via cross validation). To assess the value of learned consistency, an LC-ablated variant of DEQ-MPI was built with a conventional unlearned DC block. Training was performed at measurement SNRs of 5-45~dB, and testing was performed under 35~dB SNR. For brevity, pSNR assessments are reported, while the same conclusions are valid based on SSIM. DEQ-MPI achieves the highest performance (Fig.~\ref{fig:ablationGraph}), with pSNR improvement of 2.1~dB over the end-to-end, 3.4~dB over the unrolled, and 0.9~dB over the LC-ablated model across training SNRs. The only exception is at SNR=5~dB where the LC-ablated variant yields a moderately higher pSNR, best attributed to the relatively low training SNR that mismatches the test SNR limiting the performance of the LC block.

We also examined the convergence behaviors of the iterative models, when the training and test SNRs were both 35~dB (Fig.~\ref{fig:ablationGraphConvergence}). The non-iterative end-to-end model was not considered. The unrolled model begins to suffer dramatically when the number of iterations exceeds $\numIt$ assumed during training, and the LC-ablated variant has relatively slow convergence to a suboptimal performance level. In contrast, DEQ-MPI shows fast convergence where it exceeds the performance of the unrolled model beyond $\numIt=5$. 

Next, we assessed the importance of the proposed initializations for the RDN and LC blocks in DEQ-MPI. Accordingly, a variant based on a randomly initialized RDN, a variant based on a randomly initialized LC, and a variant with randomly initialized RDN and LC were built. To demonstrate the proposed LC initialization, an additional variant with LC pre-trained to estimate noise-free data was also built. The training and test SNRs matched (35~dB). We find that the average pSNR is 37.6~dB for DEQ-MPI, 29.9~dB when RDN is randomly initialized, 20.2~dB when LC is randomly initialized, 20.1~dB when both RDN and LC are randomly initialized, and 29.9~dB with LC pre-trained to estimate noise-free data. These results indicate that the proposed model initializations contribute substantially to reconstruction performance.

\begin{figure}[t]
\centering
     \centerline{\includegraphics[width=\columnwidth]{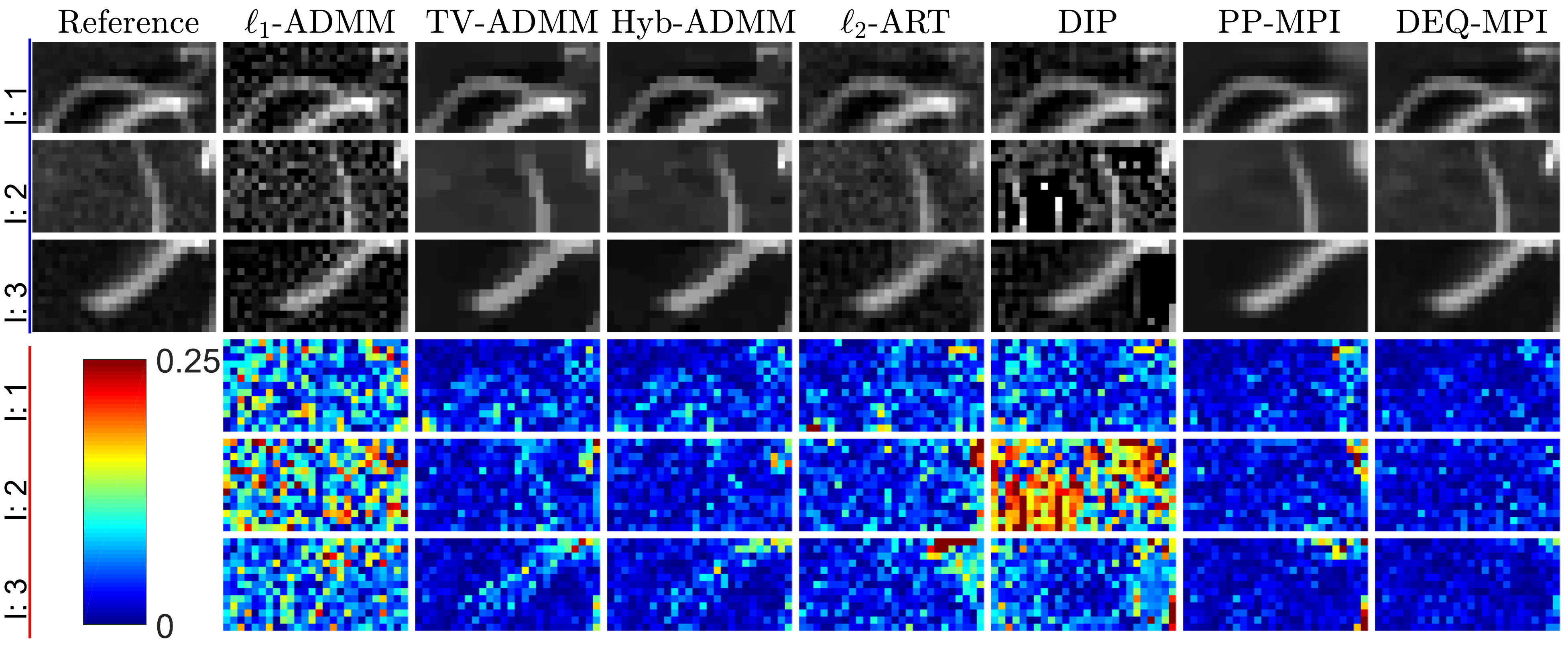}}
    \caption{Reconstructions of three simulated vessel phantoms (13$\times$26) and respective error maps (see colorbar) are shown for competing methods, along with the reference images. A measurement SNR of 35~dB was used. The SM at 0.5~T/m SF gradient was used for training, and the SM at 0.6~T/m was used for testing.}
	\label{fig:unseenByImages}
\end{figure}

\begin{figure}[t]
\centering
    \centerline{\includegraphics[width=\columnwidth]{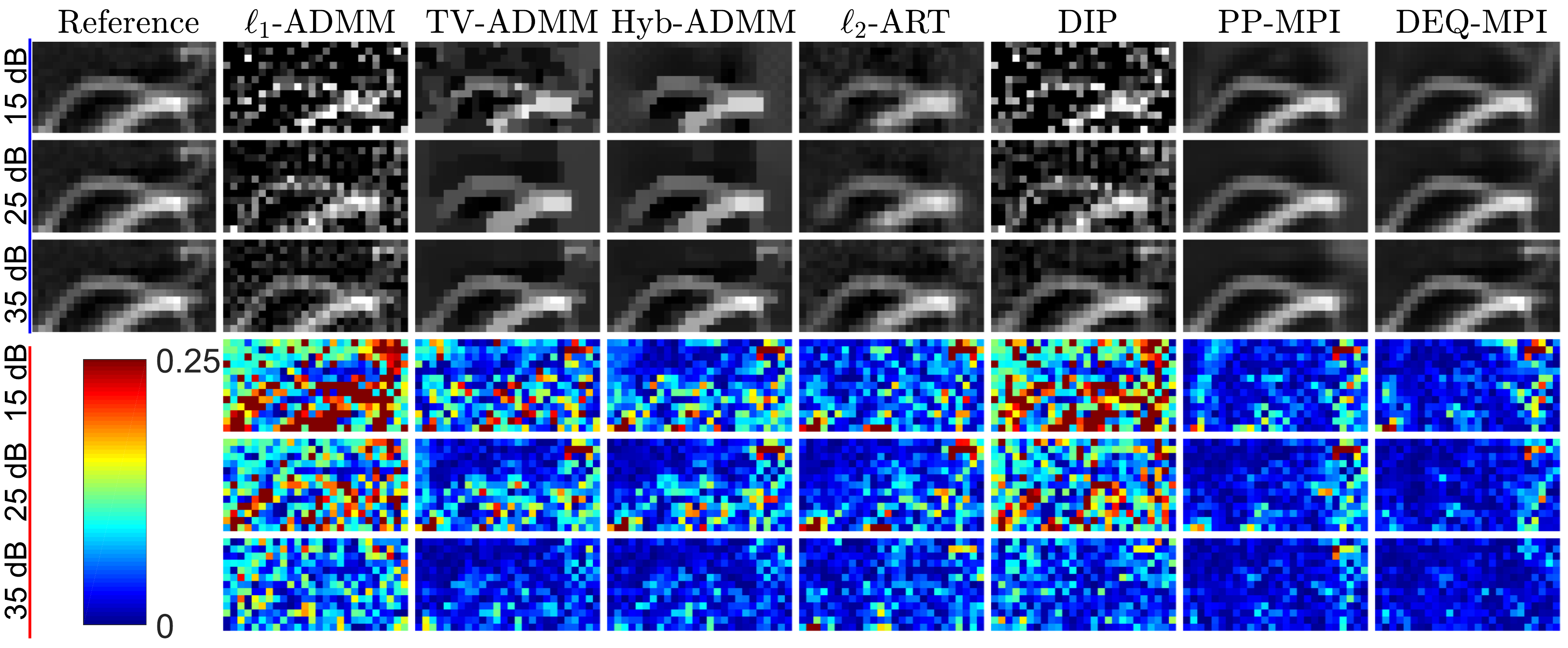}}
    \caption{Reconstructions of a simulated vessel phantom (13$\times$26) at SNR=15-35~dB and respective error maps are shown for competing methods, along with the reference image. The SM at 0.5 T/m was used for training, and the SM at 0.6 T/m was used for testing.}
	\label{fig:unseenBySNR}
\end{figure}

\begin{table}[t]
\centering
\caption{pSNR (dB)\,/\,SSIM (\%) of the simulated vessel phantoms (26$\times$52) reconstructed at varying measurement SNRs. The upsampled SM at 0.5 T/m was used in training, and the upsampled SM at 0.6 T/m was used in testing.} 
\label{table:unseenHRDatasetResults}
\setlength{\tabcolsep}{1pt}
\renewcommand{\arraystretch}{1.5}
\resizebox{1.01\columnwidth}{!}{
\begin{tabular}{|c|c|c|c|} 
\hline 
 & SNR=$15$~dB & SNR=$25$~dB & SNR=$35$~dB  \\ 
\hline
\multirow{1}{*}{$\ell_1$-ADMM} & 23.7$\pm$1.9\,/\,24.3$\pm$8.5  & 28.8$\pm$1.5\,/\,46.9$\pm$8.2  & 33.4$\pm$3.2\,/\,73.5$\pm$4.9  \\ 
 \hline 
\multirow{1}{*}{TV-ADMM} & 30.8$\pm$4.6\,/\,70.0$\pm$10.5  & 33.5$\pm$4.0\,/\,78.5$\pm$6.6  & 35.3$\pm$4.0\,/\,83.8$\pm$5.5  \\ 
 \hline 
\multirow{1}{*}{Hyb-ADMM} & 29.2$\pm$2.7\,/\,56.1$\pm$7.8  & 33.3$\pm$3.4\,/\,76.9$\pm$5.3  & 35.3$\pm$4.0\,/\,84.4$\pm$5.1  \\ 
 \hline 
\multirow{1}{*}{$\ell_2$-ART} & 28.8$\pm$4.0\,/\,55.6$\pm$5.1  & 30.3$\pm$5.4\,/\,68.7$\pm$9.3  & 33.8$\pm$4.3\,/\,78.2$\pm$5.8  \\ 
 \hline 
\multirow{1}{*}{DIP} & 23.7$\pm$1.9\,/\,25.5$\pm$11.5  & 29.3$\pm$2.3\,/\,52.1$\pm$15.2  & 31.0$\pm$3.1\,/\,64.7$\pm$16.1  \\ 
 \hline 
\multirow{1}{*}{PP-MPI} & 32.6$\pm$4.3\,/\,75.7$\pm$8.8  & 35.3$\pm$3.5\,/\,81.3$\pm$6.7  & 37.1$\pm$3.1\,/\,84.8$\pm$5.9  \\ 
 \hline 
\multirow{1}{*}{DEQ-MPI} & \textbf{33.9$\pm$3.9}\,/\,\textbf{78.7$\pm$7.8}  & \textbf{36.0$\pm$3.3}\,/\,\textbf{82.9$\pm$6.5}  & \textbf{37.7$\pm$3.2}\,/\,\textbf{86.3$\pm$5.5}  \\ 
 \hline 
\end{tabular}
}
\end{table}

\begin{figure}[t]
\centering
	\centerline{\includegraphics[width=\columnwidth]{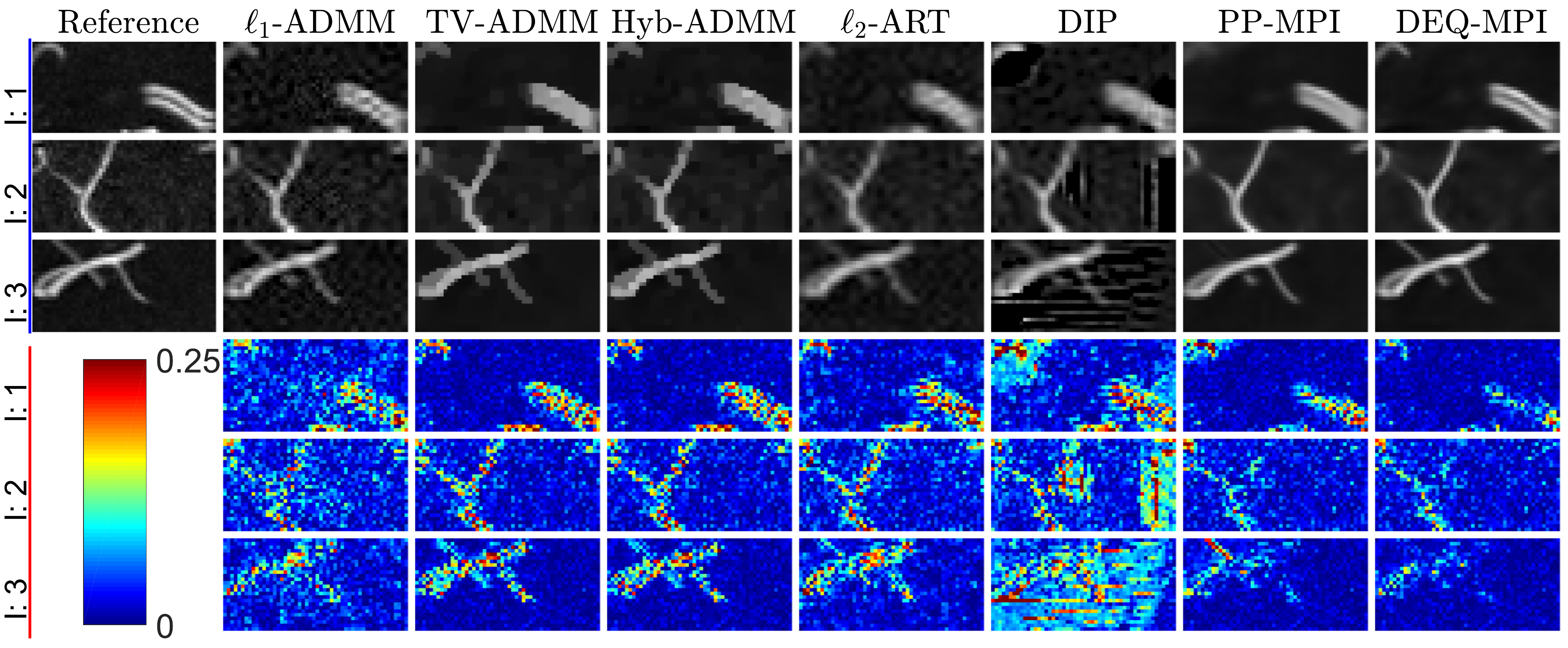}}
    \caption{Reconstructions of three simulated vessel phantoms (26$\times$52) and respective error maps are shown for competing methods, along with the reference images. A measurement SNR of 35~dB was used. The upsampled SM at 0.5~T/m was used for training, and the upsampled SM at 0.6~T/m was used for testing.}
	\label{fig:unseenHRByImages}
\end{figure}

\begin{table}[t]
\centering
\caption{pSNR (dB)\,/\,SSIM (\%) of the simulated torus-shaped phantoms (26$\times$52) reconstructed at SNR=15~dB. ID denotes inner torus diameter. The upsampled SM at 0.5 T/m was used in training, and the upsampled SM at 0.6 T/m was used in testing.} 
\label{table:unseenHRResolutionResults}
\setlength{\tabcolsep}{1pt}
\renewcommand{\arraystretch}{1.5}
\resizebox{1.01\columnwidth}{!}{
\begin{tabular}{|c|c|c|c|} 
\hline 
 & ID=$1$~mm & ID=$2$~mm & ID=$3$~mm  \\ 
\hline
\multirow{1}{*}{$\ell_1$-ADMM} & 29.6$\pm$2.2\,/\,84.5$\pm$20.4  & 29.2$\pm$2.0\,/\,86.7$\pm$19.6  & 29.0$\pm$1.9\,/\,89.5$\pm$16.8  \\ 
 \hline 
\multirow{1}{*}{TV-ADMM} & 30.0$\pm$0.9\,/\,81.7$\pm$20.1  & 28.2$\pm$1.2\,/\,82.3$\pm$17.9  & 27.7$\pm$1.6\,/\,81.9$\pm$14.6  \\ 
 \hline 
\multirow{1}{*}{Hyb-ADMM} & 30.5$\pm$1.2\,/\,96.0$\pm$3.1  & 30.2$\pm$1.2\,/\,\textbf{96.0$\pm$3.3}  & 29.8$\pm$1.2\,/\,\textbf{96.4$\pm$2.9}  \\ 
 \hline 
\multirow{1}{*}{$\ell_2$-ART} & 27.6$\pm$0.2\,/\,75.6$\pm$2.2  & 26.1$\pm$0.2\,/\,70.6$\pm$2.4  & 25.4$\pm$0.1\,/\,65.7$\pm$2.7  \\ 
 \hline 
\multirow{1}{*}{DIP} & 32.5$\pm$2.3\,/\,93.6$\pm$4.0  & 33.4$\pm$2.0\,/\,94.2$\pm$3.7  & 32.5$\pm$2.8\,/\,92.9$\pm$4.1  \\ 
 \hline 
\multirow{1}{*}{PP-MPI} & 31.6$\pm$2.3\,/\,90.8$\pm$5.0  & 31.9$\pm$2.0\,/\,89.4$\pm$4.9  & 31.6$\pm$2.0\,/\,90.3$\pm$4.9  \\ 
 \hline 
\multirow{1}{*}{DEQ-MPI} & \textbf{36.0$\pm$1.5}\,/\,\textbf{96.1$\pm$1.4}  & \textbf{35.7$\pm$1.1}\,/\,95.9$\pm$1.5  & \textbf{35.4$\pm$0.8}\,/\,95.4$\pm$1.6  \\ 
 \hline 
\end{tabular}
}
\end{table}

\begin{figure}[t]
\centering
 \centerline{\includegraphics[width=\columnwidth]{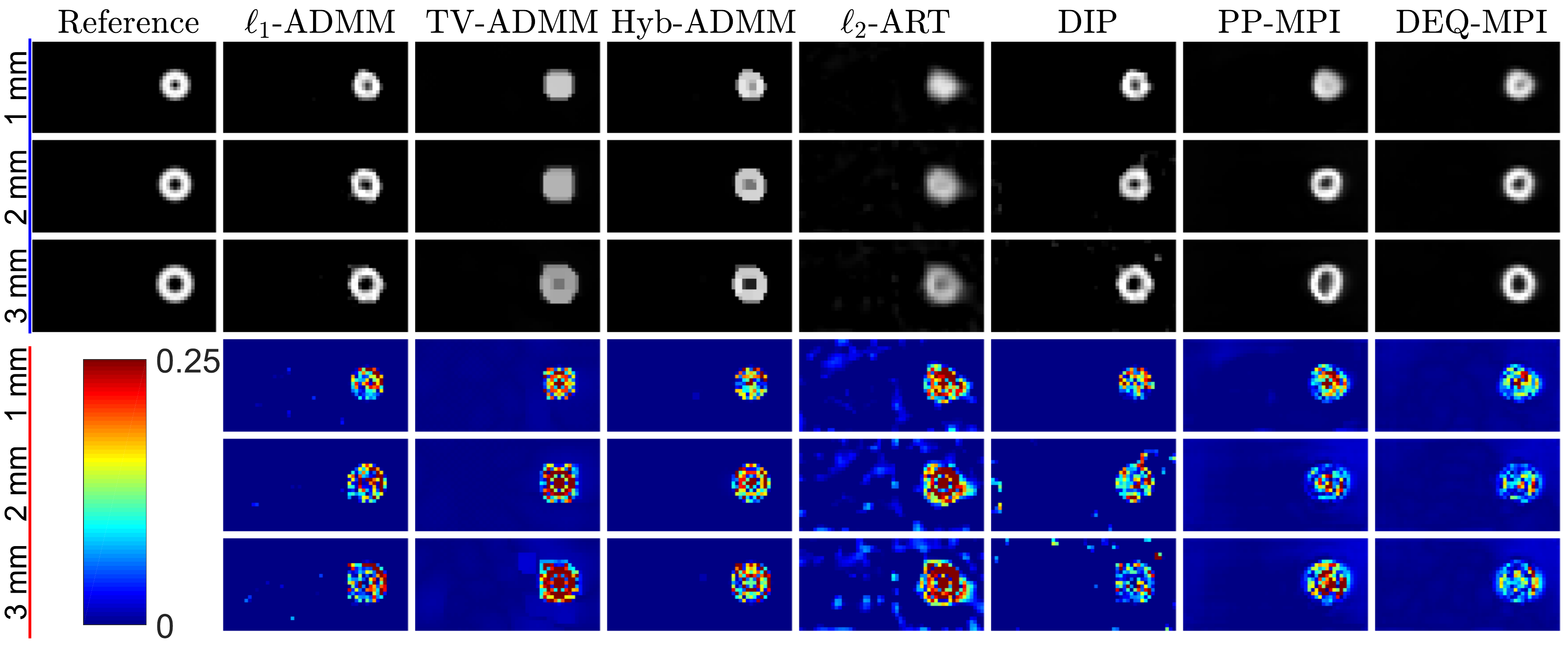}}
    \caption{Reconstructions of three torus-shaped phantoms (26$\times$52) and respective error maps with inner torus diameters of ID=1-3~mm are shown for competing methods, along with the reference images. A measurement SNR of 15~dB was used. The upsampled SM at 0.5~T/m SF gradient was used for training, and the upsampled SM at 0.6~T/m was used for testing.}
	\label{fig:unseenResolutionByImages}
\end{figure}

\vspace{-0.2cm}
\subsection{Simulated Phantoms}
DEQ-MPI was first demonstrated against traditional ($\ell_1$-ADMM, TV-ADMM, Hyb-ADMM, $\ell_2$-ART) and learning-based methods (DIP, PP-MPI) via quantitative assessments on 13$\times$26 simulated vessel phantoms. pSNR and SSIM were computed across the test set for variable measurement SNRs, with training and test sets having matching SNR for each case for DEQ-MPI. When the same SM was used for both training and testing (the SM at 0.5 T/m SF gradient from the second session), DEQ-MPI outperforms the top-contending traditional method by 2.8~dB pSNR\:/\:6.8\% SSIM, and the top-contending learning-based method by 1.7~dB pSNR\:/\:4.5\% SSIM (Table~\ref{table:seenDatasetResults}). When the SM differed across training-test sets (the SMs at 0.5 versus 0.6 T/m SF gradient), DEQ-MPI again outperforms the top-contending traditional method by 2.5~dB pSNR\:/\:6.2\% SSIM, and the top-contending learning-based method by 1.3~dB pSNR\:/\:3.7\% SSIM (Table~\ref{table:unseenDatasetResults}). For each competing method, performance levels are comparable across Tables~\ref{table:seenDatasetResults}-\ref{table:unseenDatasetResults} because both cases utilized the same underlying phantoms in the test set and the same SNR levels. Note, however, that this does not imply that the SMs at 0.5 and 0.6 T/m are interchangeable, since reconstructing the data measured with the SM at 0.5 T/m using the SM at 0.6 T/m would result in substantial performance loss (e.g. 15.5~dB pSNR\:/\:7.3\% SSIM loss for Hyb-ADMM).

\begin{figure*}[t]
    \centering
	\includegraphics[width=0.77\textwidth]{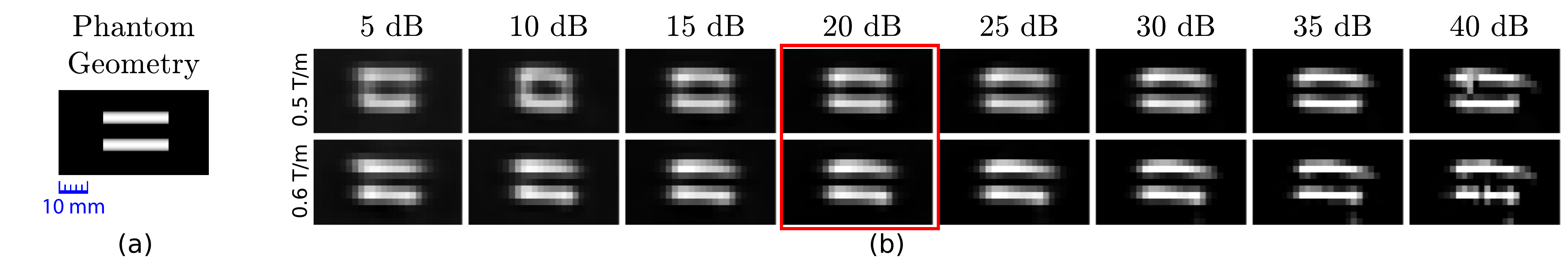}
    \caption{Reconstructions of the experimental cylindrical phantom with DEQ-MPI trained at SNR levels in 5-40~dB. \textbf{(a)} Approximate geometry of the phantom. Scale bar indicates 10 mm. \textbf{(b)} Reconstructions at \textbf{(top-row)} 0.5 T/m and \textbf{(bottom-row)} 0.6 T/m SF gradients. The SM at 0.5 T/m was used for training DEQ-MPI. Testing utilized the SM corresponding to each case. Red box denotes the case where the training and testing SNRs match.}
	\label{fig:CPhantomSNR}
\end{figure*}

\begin{figure}[t]
    \centering
\centerline{\includegraphics[width=0.55\columnwidth]{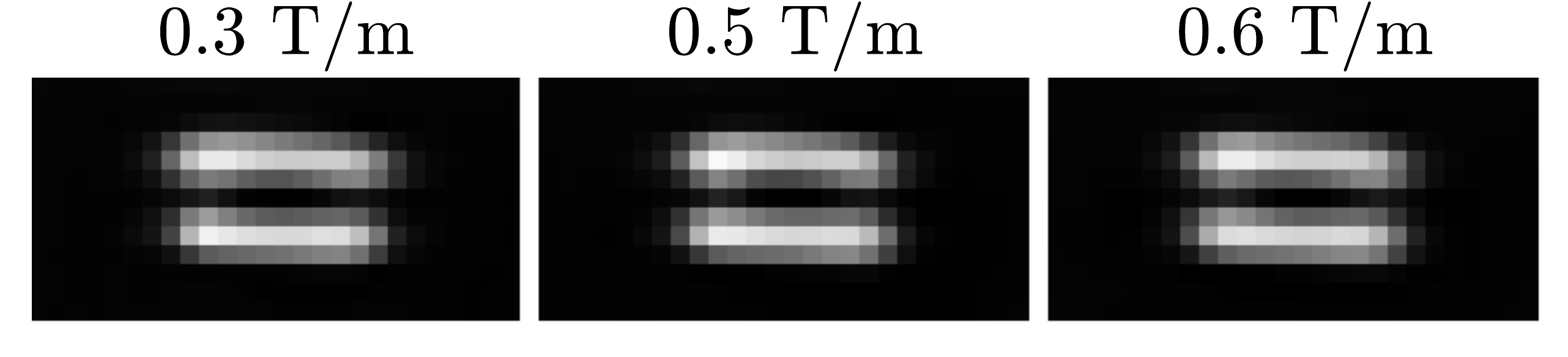}}
    \caption{Reconstructions of the experimental cylindrical phantom with DEQ-MPI trained at SF gradients of 0.3, 0.5, or 0.6 T/m. In all cases, testing utilized the SM at 0.5~T/m.}
	\label{fig:SFExperiment}
\end{figure}

Representative reconstructions and the respective error maps from competing methods under $35$ dB SNR are displayed in Fig. \ref{fig:unseenByImages} for the case with different SMs across the training-test sets. $\ell_1$-ADMM yields a grainy image with residual noise and background signal; TV-ADMM, Hyb-ADMM, and $\ell_2$-ART suffer from spatial blurring; and DIP can suffer from noise amplification. While PP-MPI yields relatively higher performance, it shows elevated errors in regions of low signal near the upper and lower right corners, where the experimental SM has limited sensitivity due to the limits of receive coil coverage. In contrast, DEQ-MPI yields superior performance with lower errors than competing methods. Reconstructions for varying measurement SNRs for the case with different SMs across the training-test sets are shown in Fig.~\ref{fig:unseenBySNR}. As expected, performance improves for all methods as measurement SNR increases. Among competing methods, $\ell_1$-ADMM and particularly DIP that are relatively amenable to noise amplification show limited performance towards lower SNR levels. Overall, DEQ-MPI produces high image quality with lower artifacts and noise than competing methods. 

Demonstrations were also performed on simulated vessel phantoms at a larger grid size of 26$\times$52. To account for the larger grid size, the measured SMs were upsampled via bicubic interpolation to 1 mm/pixel resolution. Performance was quantified for variable measurement SNRs, while the SMs differed between the training and test sets. DEQ-MPI outperforms the top-contending traditional method by 2.7~dB pSNR\:/\:5.2\% SSIM, and the top-contending learning-based method by 0.9~dB pSNR\:/\:2.0\% SSIM (Table~\ref{table:unseenHRDatasetResults}). Representative reconstructions under $35$ dB SNR are displayed in Fig. \ref{fig:unseenHRByImages}. Among the competing methods, $\ell_1$-ADMM and DIP show noise amplification, TV-ADMM and Hyb-ADMM show block artifacts, and $\ell_2$-ART shows blurring. In contrast, DEQ-MPI recovers images with higher spatial acuity and lower errors than competing methods. 

To systematically assess resolvability of fine structure, DEQ-MPI was demonstrated using 26$\times$52 simulated torus-shaped phantoms. Separate phantoms were generated with an MNP-free torus diameter gradually reduced from 3-mm to 1-mm. Performance was quantified for 15 dB measurement SNR and mismatched SMs between the training and test sets. DEQ-MPI outperforms the top-contending traditional method by 5.5~dB pSNR while offering similar SSIM, and the top-contending learning-based method by 2.9~dB pSNR\:/\:2.3\% SSIM (Table~\ref{table:unseenHRResolutionResults}). Note that the phantoms examined in this analysis are highly sparse with MNPs located only within a small torus. pSNR is based on absolute pixel-wise errors without any local normalization, whereas SSIM is based on relative window-wise similarities with window-level normalization. As such, pSNR values remain more sensitive to errors near the torus region, while SSIM values are dominated by the close match between reconstructed and reference images in void background regions. Hence, we deduce that pSNR better reflects the performances of the competing methods in this case. Representative reconstructed phantom images are shown in Fig.~\ref{fig:unseenResolutionByImages}. The torus-shaped phantom was placed off-centered within the FOV to present a more challenging case for all methods, as the measured SMs had reduced sensitivity in the peripheries of the FOV. Among competing methods, TV-ADMM and Hyb-ADMM show blocking artifacts, $\ell_2$-ART shows blurring, and DIP shows pixel artifacts due to noise amplification that limit spatial acuity. Although $\ell_1$-ADMM yields visually sharp reconstructions, close inspection of reconstructed images reveals that it suffers from amplitude errors due to undershooting or overshooting of pixel intensities. Meanwhile, PP-MPI yields relatively lower artifacts, but it shows geometric distortions in the recovered torus, particularly visible at larger inner diameters. In comparison, DEQ-MPI recovers the torus shape with minimal artifacts and distortions, and successfully resolves the reduced intensity in the MNP-free inner region for inner torus diameter as low as 1 mm.

\begin{figure*}
    \centering
\centerline{\includegraphics[width=0.70\textwidth]{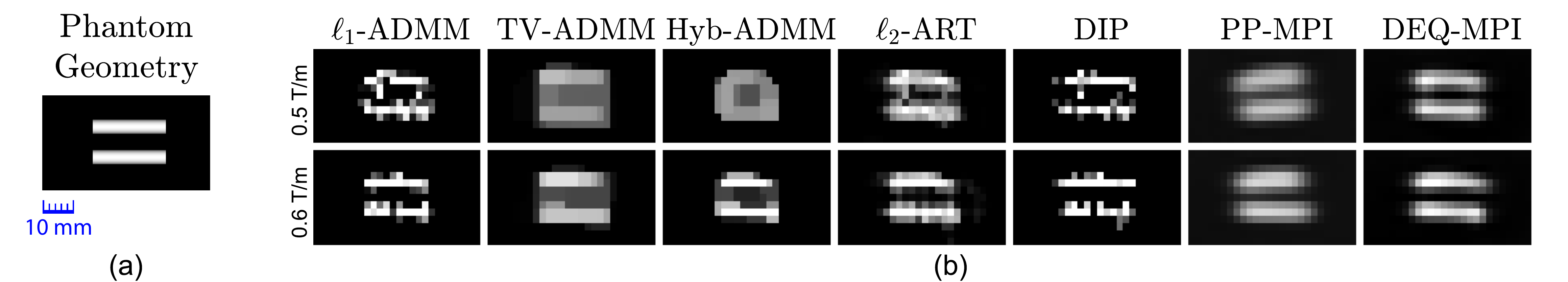}}
    \caption{Reconstructions of experimental cylindrical phantom with competing methods. \textbf{(a)} Approximate geometry of the phantom. \textbf{(b)} Reconstructed images at \textbf{(top-row)} 0.5 T/m and \textbf{(bottom-row)} 0.6 T/m SF gradients. The SM at 0.5 T/m was used for training DEQ-MPI. Testing utilized the SM corresponding to each case.
    }
	\label{fig:CPhantom}
\end{figure*}

\begin{figure*}
\centering
\centerline{\includegraphics[width=0.70\textwidth]{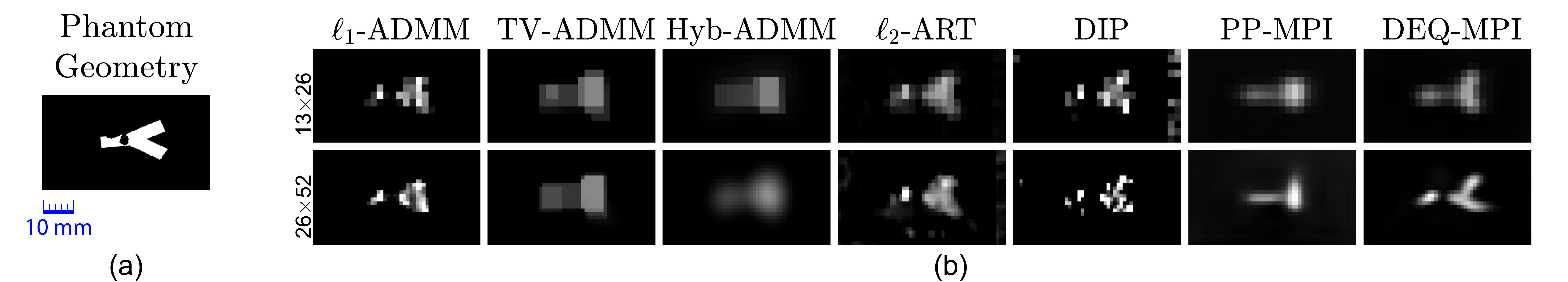}}
\caption{Reconstructions of experimental Y-shape phantom with competing methods. \textbf{(a)} Approximate geometry of the phantom. \textbf{(b)} Reconstructed images based on \textbf{(top-row)} the original $13 \times 26$ SM and \textbf{(bottom-row)} the upsampled $26\times 52$ SM. The SM at 0.5 T/m from the second session was used for training DEQ-MPI. The SM at 0.5 T/m from the first session was used for testing.}
\label{fig:YPhantom}
\end{figure*}

\subsection{Experimental Phantoms}
Next, DEQ-MPI was demonstrated on two experimental phantoms. Due to lack of ground-truth images in experimental settings, assessments were performed qualitatively via visual inspection \cite{Storath2017, cbtop2020}. DEQ-MPI was trained using emulated MPI data from simulated phantoms at an assumed SNR level, whereas testing was performed on experimental MPI data at SNR$\approx$20~dB (estimated based on multiple signal and background measurements). To assess reliability against SNR and SM mismatches between the training-test sets, reconstructions of measurements at 0.5 and 0.6 T/m SF gradients were obtained separately using models independently trained for SNRs in 5-40~dB (Fig.~\ref{fig:CPhantomSNR}). Training was performed using the SM at 0.5 T/m, whereas testing was performed using the SM corresponding to each SF gradient. In general, DEQ-MPI shows reliability against moderate deviations between the training and test SNRs, albeit residual reconstruction errors occur when the difference between the two SNRs reaches towards 20~dB. In particular, residual artifacts become apparent in reconstructed images when the model trained at 40 dB SNR is tested at 20 dB SNR. This finding is best attributed to the large mismatch between the training and test SNR levels, which can limit generalization and cause over-sensitivity to noise. To further assess reliability against SM mismatches between the training-test sets, separate reconstructions of the cylindrical phantom measurement at 0.5 T/m SF gradient were obtained using models trained separately with SMs at SF gradients of 0.3, 0.5 or 0.6 T/m (Fig.~\ref{fig:SFExperiment}). We observe minimal differences in reconstructions for models trained at 0.3-0.6 T/m, suggesting that DEQ-MPI demonstrates a degree of robustness against SM deviations.

Reconstructions of the cylindrical phantom were then compared for the competing methods at 0.5 and 0.6 T/m (Fig.~\ref{fig:CPhantom}). Again, training for DEQ-MPI was performed using the SM at 0.5 T/m, whereas testing was performed using the SM corresponding to each case. The other methods utilized the SM corresponding to each case, as well. $\ell_1$-ADMM and DIP yield over-sparsified images with artefactual bright/dark pixels due to amplified noise; TV-ADMM and Hyb-ADMM yield over-smoothed images with block artifacts; and $\ell_2$-ART shows blurring and residual noise. While PP-MPI mostly avoids these issues, it reconstructs cylindrical tubes at an incorrect geometric orientation compared to remaining methods. In contrast, DEQ-MPI yields lower artifacts/noise and higher resemblance to the designed phantom than competing methods.   

Images of the Y-shape phantom were also reconstructed (Fig.~\ref{fig:YPhantom}). The SM at 0.5 T/m from the second experimental session was used for training DEQ-MPI, whereas testing was performed using the SM at 0.5 T/m from the first session. A recalibration was performed on the FFL system between the two sessions that were 2 months apart, so the resultant SMs differed. The measured SM (sampled at 2-mm/pixel) was used to reconstruct images at the original $13\times26$ grid size, and bicubic interpolated version of the SM (upsampled to 1-mm/pixel) was used to reconstruct images at $26\times52$ grid size \cite{Gungor_iwmpi, cbtop2020}. $\ell_1$-ADMM and DIP suffer from artefactual pixels due to noise amplification; TV-ADMM, Hyb-ADMM, and $\ell_2$-ART suffer from spatial blurring; and PP-MPI does not faithfully capture the geometry of the phantom including the central air bubble. In comparison, DEQ-MPI offers high quality reconstructions in both the original and upsampled resolutions.

\subsection{Reconstruction Time}
\begin{figure}[t]
\centering
\centerline{\includegraphics[width=0.95\columnwidth]{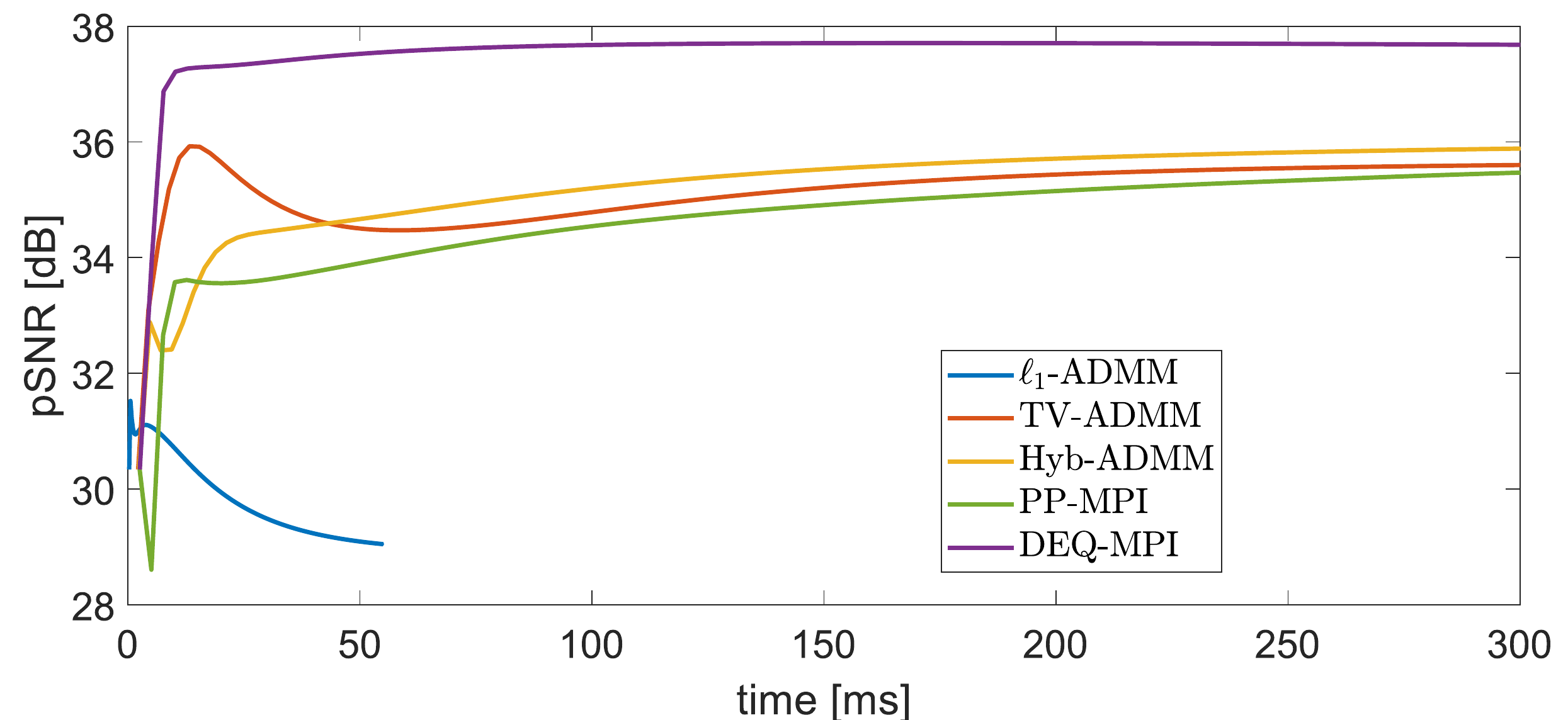}}
\caption{Performance of competing methods as a function of inference time. $\ell_2$-ART and DIP that have markedly prolonged run times are omitted. Average pSNR across the test set is shown for each method.}
	\label{fig:unseenConvergenceTest}
\end{figure}

The number of inference iterations and reconstruction times for all competing methods are listed in Table~\ref{table:reconTime}. Reconstruction performance as a function of run time is plotted in Fig. \ref{fig:unseenConvergenceTest}. Among competing methods, $\ell_2$-ART and particularly DIP require prolonged inference, and ADMM variants with TV regularization (TV-ADMM, Hyb-ADMM) and PP-MPI have moderate run times. In comparison, DEQ-MPI yields efficient reconstructions with relatively fast convergence and run times competitive with $\ell_1$-ADMM.

\begin{table}[t]
\centering
\caption{Inference iterations (number of iterations) and reconstruction times (milliseconds) for a single 13$\times$26 MPI image.} 
\label{table:reconTime}
\setlength{\tabcolsep}{2pt}
\resizebox{1.01\columnwidth}{!}{
\begin{tabular}{|c|c|c|c|c|c|c|c|} 
\hline 
 & $\ell_1$-ADMM & TV-ADMM & Hyb-ADMM & $\ell_2$-ART & DIP & PP-MPI & DEQ-MPI \\ 
\hline
Iters. & 200 & 100 & 100 & 10 & 20000 & 150 & 25 \\
\hline
Time & 55 & 221 & 236 & 6966 & 235162 & 379 & 64 \\
\hline
\end{tabular}}
\end{table}

\section{Discussion}
DEQ-MPI integrates an implicit mapping into an optimization algorithm for performance and efficiency in MPI reconstruction. The implicit mapping is based on learned regularization and DC blocks to better conform to the data distribution, and accelerated fixed-point iterations are used to rapidly compute a convergent solution. Demonstrations on simulated and experimental phantoms indicate that DEQ-MPI trained using a single acquired SM outperforms previous traditional and learning-based methods. As a physics-driven method, DEQ-MPI shows reliability against deviations in the SM and in SNR levels between the training and test sets. While reconstruction errors occur when the training SNR is dramatically higher than the test SNR, this scenario can be avoided by performing a rough SNR estimation on given data. 

MPI reconstruction involves the solution of an ill-conditioned inverse problem due to significant measurement correlations and high noise levels. Ill-conditioning can notably degrade image quality, and in turn restrict the use of MNPs with suboptimal characteristics (e.g., spatially-broad PSF, weak responses at high harmonic frequency components). Physics-driven deep learning methods integrate data-driven image priors with physical constraints of the imaging system to effectively regularize reconstructions, while maintaining reasonable robustness against changes in the system constraints \cite{kamilovPnP2,Lam2023}. By enhancing image quality over traditional reconstructions, physics-driven methods such as DEQ-MPI can enable high-performance imaging even when utilizing MNPs with less desirable characteristics. Future studies are warranted to systematically assess the utility of DEQ-MPI in enabling use of a broader variety of MNPs in MPI.

The SMs acquired on our in-house MPI scanner had a resolution of 2 mm/pixel given limitations related to the SF gradient strength and MNP characteristics, as typically encountered in MPI systems \cite{saritas2012rev}. For assessments at 1-mm/pixel resolution, the measured SMs were upsampled via bicubic interpolation and  26$\times$52 images were reconstructed. Analyses on simulated vessel and torus-shaped phantoms suggest that DEQ-MPI can faithfully reconstruct features at spatial scales down to 1 mm. Yet, the ability to resolve fine features depends on various critical factors beyond the reconstruction method, including the compatibility between the upsampled SM and the actual high-resolution SM, distribution of the singular values of the SM, measurement SNR, and position of the MNP sample within the FOV. Thus, future studies are warranted to experimentally investigate the ability of DEQ-MPI in resolving features below 1-mm scale by measuring higher-resolution SMs, and the benefits of DEQ-MPI over competing methods in recovering images of larger sizes.

Traditional methods can show high sensitivity to weights for hand-crafted regularizers \cite{Dar2020}. MPI studies have reported that ideal weights can vary drastically across scans, suggesting that parameter tuning on each test image might be useful \cite{cbtop2020, Storath2017}. Such optimization is infeasible in pre-clinical or clinical scenarios as no a priori knowledge would be available on the MNP distribution. To address this challenge, here we optimized model hyperparameters on a validation set, and the selected values were used thereafter in the test set. Learning-based methods were observed to be more forgiving against suboptimal parameters (results not shown), so they might alleviate the need for exhaustive parameter tuning.

While performant reconstructions have been reported based on untrained networks in MPI literature \cite{deepImagePriorMPI,knopp2022warmstart}, here we observed relatively limited performance with DIP. Note that DIP directly minimizes a conventional DC loss between recovered and acquired test data. This loss function intrinsically assumes that data contain negligible noise compared to the signal. Because this assumption is violated for moderate to low SNR levels as considered in the current study, DIP can perform suboptimally in relatively limited SNR regimes. 

Few recent studies have considered deep equilibrium models for undersampled MRI reconstruction \cite{dongLiangDEQPOCS,willett, deqMeasurementModel}, and low-dose CT reconstruction \cite{Heaton2021}. In addition to addressing a distinct problem in MPI, our proposed approach is unique in the following aspects: (1) Instead of integrating an implicit mapping into a projection-onto-convex-sets algorithm as in \cite{dongLiangDEQPOCS} or into a proximal gradient algorithm as in  \cite{deqMeasurementModel}, DEQ-MPI leverages an ADMM algorithm that can offer improved reliability for non-convex or non-smooth problems. (2) While \cite{willett} uses ADMM with Anderson acceleration similar to the proposed method, the two methods differ in their variable splitting procedures for ADMM. \cite{willett} uses a single-component auxiliary variable dedicated to the proximal mapping for regularization. In contrast, DEQ-MPI leverages a two-component auxiliary variable with sub-components dedicated to the proximal mappings for data consistency and regularization, respectively. In initial phases of the study, we observed that this splitting procedure facilitates implementation of a constrained ADMM formulation based on a learned consistency measure. (3) Unlike \cite{Heaton2021} that uses Jacobian-free backpropagation, DEQ-MPI employs implicit differentiation for model training. (4) \cite{dongLiangDEQPOCS,willett, deqMeasurementModel, Heaton2021} all employ a conventional unlearned DC block. In contrast, DEQ-MPI leverages a learned consistency (LC) block to better conform to the data distribution. (5) While \cite{dongLiangDEQPOCS,Heaton2021} do not report non-standard initialization and \cite{willett,deqMeasurementModel} only consider initialization for the regularization block, DEQ-MPI employs dedicated initialization methods for its regularization and LC blocks that improve model performance. (6) Lastly, \cite{dongLiangDEQPOCS, willett, deqMeasurementModel} use a residual connection between the input and output layers of a convolutional architecture, and \cite{Heaton2021} uses four residual convolutional blocks with a residual connection between the input and output of each block. Instead, DEQ-MPI adopts multiple residual connections densely distributed across layers in a convolutional architecture that have been reported to offer performance benefits \cite{rdn}. 

Several developments can be considered to improve DEQ-MPI. First, we generated training data using MRA images under the assumption that they have similar features to MPI images. When the imaged anatomy is non-vascular, this approach might yield suboptimal performance. While public datasets of MPI images are rare, DEQ-MPI can in principle be enhanced by training the model on large amounts of experimental data to better capture application-specific image features. Second, MPI data include a non-stationary background that was separately measured and subtracted from acquired data prior to reconstruction. The need for background measurements can be avoided by extending DEQ-MPI to separately reconstruct the foreground and background signals. To do this, a dictionary-based approach can be adopted to estimate the background signal from acquired data \cite{Knopp_2021_background}. Third, DEQ-MPI was trained based on convolutional networks and a pixel-wise loss term. Performance improvements might be achieved with attention-based architectures to capture contextual features \cite{resvit}, and diffusion processes to increase reliability in model training \cite{syndiff}. Fourth, DEQ-MPI was demonstrated for reconstructing experimental phantoms at 2$\times$ higher spatial resolution by bicubic SM upsampling. Visual acuity of resultant MPI images 
scales up well, suggesting that upsampled SMs are reasonably accurate. To enhance accuracy, learning-based super-resolution methods can also be adopted \cite{2d-SMRnet,transmsTMI}. It remains an important future work to evaluate the utility of upsampling methods via comparisons against SMs acquired at high resolution. Fifth, DEQ-MPI was trained via common backpropagation, where the Jacobian of the convergent solution was computed via implicit differentiation. A powerful alternative is the Jacobian-free backpropagation framework that improves training efficiency and numerical stability \cite{WuFung2022JFB,Heaton2022learn}, which can be utilized in DEQ-MPI to lower training costs and enhance reconstruction performance. Finally, imaging over large FOVs can be attained by performing patch-wise reconstructions with DEQ-MPI and fusing the multi-patch outputs \cite{multiPatchMPI}.

\section{Conclusion}
Here, we introduced a novel deep equilibrium reconstruction for MPI with learned consistency. For improved performance and reliability, DEQ-MPI follows a physics-driven approach that integrates network blocks that regularize the image and enforce DC into an iterative algorithm. Simulated and experimental demonstrations indicate clear performance benefits and competitive efficiency over both traditional and recent learning-based methods. Thus, DEQ-MPI holds great promise for fast, high-fidelity image reconstruction in MPI.

\bibliographystyle{IEEEtran}
\bibliography{IEEEabrv,refs}

\end{document}